\documentclass[lettersize,journal]{IEEEtran}
\usepackage{amsmath,amsfonts}
\usepackage{algorithmic}
\usepackage{algorithm}
\usepackage{array}
\usepackage[caption=false,font=normalsize,labelfont=sf,textfont=sf]{subfig}
\usepackage{textcomp}
\usepackage{stfloats}
\usepackage{url}
\usepackage{verbatim}
\usepackage{graphicx}
\usepackage{cite}
\begin{document}

\title{Design, Modeling, and Validation of Curvature-Based Shape-Aware Flexible Phased Arrays}

\author{Yair Dashevsky~\IEEEmembership{Student Member,~IEEE,}
        Alon Elgarat~\IEEEmembership{Student Member,~IEEE,}
        Matan Gal-Katziri,~\IEEEmembership{Member,~IEEE}

\thanks{This work was supported by the Israeli Science Foundation (ISF) No. 1234/24.}

\thanks{The authors are with Ben-Gurion University, School of Electrical and Computer Engineering, Beer-Sheva, Israel (e-mail: yairdash@post.bgu.ac.il).}}



\maketitle

\begin{abstract}
In situ knowledge of array geometry enables mechanically flexible phased arrays to recover beam-steering performance under deformation without receiver-side RF feedback. This work presents a comprehensive engineering study of curvature-based shape sensing, combining numerical sampling-density and measurement-noise analysis, physical modeling of sensor response, and experimental shape and RF characterization. A study of 11,000 B\'ezier curves quantifies reconstruction error versus curvature-sampling density and noise, guiding the selection of nine sensing locations. A physical model relates PCB curvature to strain-gauge output voltage, with predictions compared against calibrated sensor measurements. The sensing system is integrated with an eight-element, 6-GHz flexible array and a low-cost, size-scalable, digitally controlled eight-channel phase-shifting module providing 360° tuning with better than 7-bit effective resolution. Across 12 fixture-conformed shapes, including five previously reported cases, the average reconstruction error is 0.06$\lambda$. Three additional air-suspended configurations, not prescribed by analytical curves, show an average positional difference of 0.095$\lambda$ between curvature-based and image-based estimates. Measured radiation patterns demonstrate broadside correction and steering to ±30°. These results provide quantitative engineering guidance for designing and characterizing curvature-based shape-aware flexible phased arrays.
\end{abstract}

\section{Introduction}
\label{intro}
\IEEEPARstart{M}{echanically} flexible phased arrays, along with other shape-changing antenna systems \cite{suresh2022origami_sengupta,Jamal2025Origami,austin2022jom_flex, Sun2024metasurface}, represent an emerging direction in smart antenna technology. Enabled by progress in miniaturization, algorithm design, and materials engineering, such arrays are designed to be lightweight and conformal while maintaining a low-profile form factor. These properties extend the use of phased arrays to domains that were previously impractical, including reconfigurable reflecting surfaces \cite{hu2022tile_ims_manos, Rao2022reflective_surface}, body-worn antennas \cite{he2021inkjet}, low-mass spaceborne systems \cite{you2023lcp_okada, Chen2012conformal_2d}, modular deployable apertures \cite{matan2022npj}, and future ultralight autonomous platforms \cite{rafsanjani2018kirigami_robot}. Like other phased arrays, mechanically flexible arrays require calibration to mitigate gain and phase or delay errors caused by RF chain mismatches, fabrication tolerances, and interelement coupling.
\begin{figure}[t]
    \centering
    \includegraphics[clip, trim=0cm 0.1cm 0cm 0cm,width=\linewidth]{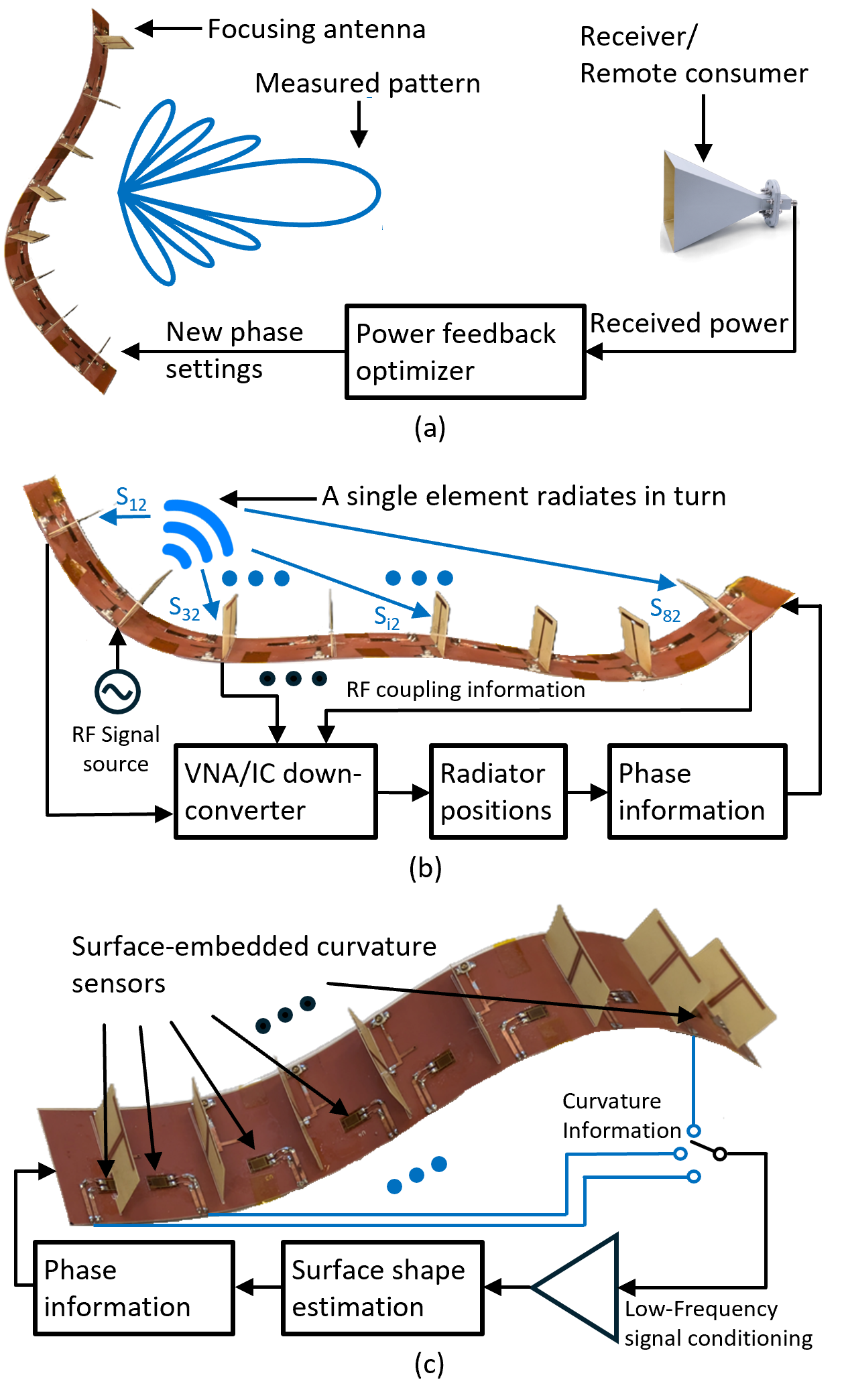}
    \caption{Methods for mitigating deformation effects in flexible phased arrays: (a) feedback optimization, (b) shape reconstruction from mutual coupling between radiators, and (c) direct surface-curvature measurement.}
    \label{fig:3_reconstruction_methods}
\end{figure}
In flexible arrays, however, mechanical deformation additionally perturbs element spacing, orientations, and feed paths, often in a time-varying manner, thereby changing the signal and radiation characteristics. These effects must be compensated in situ and in real time to maintain performance. Two fundamentally distinct approaches have been proposed to address this challenge. The first, illustrated in Fig. \ref{fig:3_reconstruction_methods}a, employs feedback optimization to identify the array excitations required to direct a signal toward a specified location. Using optimization techniques \cite{ali2020jssc,Poolakkal2025flex_nature_comm} or learning algorithms \cite{Zhou2023ml_active_array,zardi2021ml_adaptive_array,zhang2022ml_conformal,Kim2023dnn_ml_opa}, this approach abstracts radiated-field uncertainties into the learning or optimization process,
%
%
including those arising from fabrication tolerances, mutual coupling, element placement, and even nonlinear behavior. However, despite its strengths, this method has several limitations in practical, dynamic environments. Calibration time may not scale favorably for large arrays with thousands or even millions of elements \cite{Clarck2010million_element_array, sspp2020website}; feedback latency, especially under significant interference or over long TX–RX distances, further slows convergence; and optimizing over a finite set of spatial probes without incorporating information about the array’s physical shape can yield a suboptimal beam pattern and elevated sidelobe levels.
The second approach seeks to estimate the unknown array shape to derive the information needed to mitigate deformation effects. This can be done, for example, by measuring the mutual coupling between array radiators \cite{ShapeCal} as illustrated in Fig. \ref{fig:3_reconstruction_methods}b. Leveraging transceivers already integrated into the array, this method requires neither external feedback nor additional hardware, making it suitable for single-node applications such as standalone environmental sensors. Nonetheless, electromagnetic coupling is sensitive to changes in the surrounding medium, which are not easily distinguishable from changes in element positions in the measured response. Multipath reflections, potential surface-waves, and other interference degrade the signal-to-noise ratio (SNR) in coupling measurements, thereby necessitating substantial sampling redundancy for accurate shape estimation. In current implementations, using the same radio for both data transfer and shape estimation also requires halting normal system operation during measurement, limiting its applicability in dynamic scenarios.
Another modality of the shape-reconstruction approach is based on direct measurement of the surface itself. Methods that sample relative positions \cite{Molleda2011laser_shapecal} or rely on optical imaging \cite{Talon2022optical_shapecal,Joshi2024optical_Shapecal} have been demonstrated, but require bulky/external equipment, which undermines system flexibility and limits the minimum achievable form factor. Alternatively, indirect measurements such as surface curvature or relative orientation can be used to reconstruct the array’s shape from local strain \cite{yair2025shapecal} or gyroscope \cite{Zhi2025Gyroscopes} data; until recently, however, this approach had been demonstrated only on simple geometries \cite{braaten2013selflex}, on arrays with predetermined folding locations \cite{braaten2014selflex_conference, Rigobello2017strongdef}, or on structures with small curvatures \cite{zhou2019adaptivecomp}.

This work expands on our recently published generalized shape-estimation method \cite{yair2025shapecal} (Fig. \ref{fig:3_reconstruction_methods}c), which reconstructs the shape of flexible phased arrays from local curvature measurements. As previously demonstrated, the proposed solution significantly improves upon prior methods in handling unknown, complex, and highly curved shapes. In this article, we present a sampling-density and performance analysis for determining the sensor placement required to achieve the desired reconstruction accuracy under practical physical constraints. We also detail the physical modeling, design, and characterization of the curvature-sensing mechanism, showing how commercially available strain gauges can be integrated into flexible arrays with predictable output response. In addition, we present the design and construction of an eight-channel, high-resolution, low-cost, and size-scalable digitally controlled phase shifter, intended as a key experimental platform for characterizing smart antennas. The design is useful for a wide range of related studies in which the cost and complexity of increasing the number of active antennas may otherwise be limiting factors. Finally, we extend the experimental validation of the method to additional conformed and free-hanging test cases, demonstrating both shape reconstruction and recovery of beam-steering performance under deformation.
\begin{figure}
\centering\includegraphics[clip, trim=0cm 0cm 0cm 0cm, width=\linewidth]{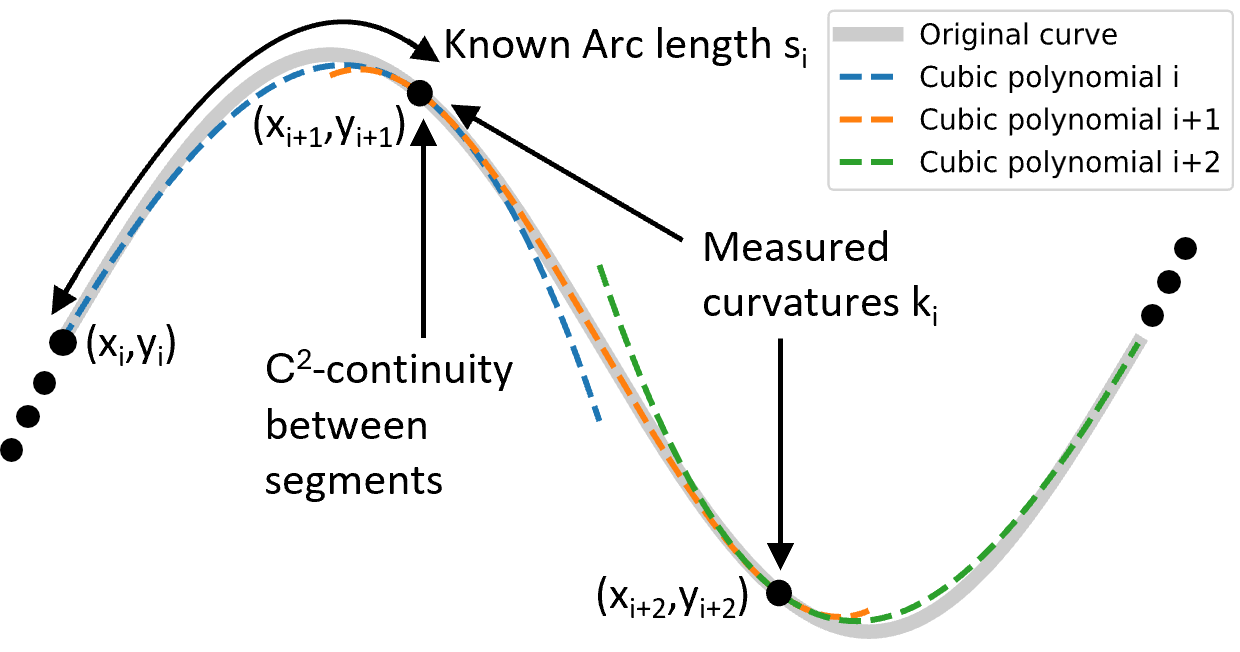}\caption{Shape estimation from curvature measurements in 1-D (illustration).}
\label{fig:curve_estim}
\end{figure}

\section{Shape Estimation From Curvature Measurements}
\label{sec:SHAPE ESTIMATION FROM CURVATURE MEASUREMENTS}

The outline for shape estimation from curvature measurements was introduced in \cite{yair2025shapecal}, wherein a sequence of interconnected cubic polynomials was used to approximate the unknown geometry of a 1-D curve. The process is similar to spline interpolation \cite{spath1974spline}, except that in our case the positions of the sample points are unknown. Instead, the polynomials are constrained by the measured curvature at each sample point and by the arc lengths between them which are known design parameters. As illustrated in Fig. \ref{fig:curve_estim}, using $N$ sample points $n_i=[x_i,y_i]$, $i=0,\ldots,N-1$, enables approximation of an unknown curve with $N-1$ cubic segments of the form
$y_i(x)=a_i x^3 + b_i x^2 + c_i x + d_i$, where $i=1,\ldots,N-1$.
The known arc lengths and the measured curvatures at the sample points, together with continuity constraints, provide sufficient data to determine the sensor positions and the polynomial coefficients by solving (at the sample points)
\begin{equation}
\label{curv}
\frac{y_i''}{\left(1+y_i'^2\right)^{3/2}} = k_i,
\end{equation}
\begin{equation}
\label{arc}
\int{ds_i}=s_i,
\end{equation}
\begin{equation}
\label{cont}
y_i=y_{i+1},
\end{equation}
\begin{equation}
\label{slope_cont}
y'_i=y'_{i+1},
\end{equation}
\begin{equation}
\label{second_deriv_cont}
y''_i=y''_{i+1}.
\end{equation}

Here, $y_i'$ and $y_i''$ denote the first and second derivatives with respect to $x$ of the $i$-th polynomial $y_i$; $k_i$ is the curvature at the $i$-th sample point; $s_i$ is the arc length between segments $i$ and $i+1$; and $ds_i$ is the differential element of arc length along the $i$-th segment, whose integral yields $s_i$ between sample points $i$ and $i+1$. Equations \eqref{curv} and \eqref{arc} correspond to the measured curvature at the designated sample point and the known arc length between sample points, respectively. Equations \eqref{cont}, \eqref{slope_cont}, and \eqref{second_deriv_cont} enforce continuity by constraining the function value, slope, and second derivative of adjacent segments to be equal at their intersection points. Notably, this formulation slightly differs from \cite{yair2025shapecal}, which enforced equal curvatures at the intersections of adjacent segments rather than equal second derivatives. This change was adopted because curvature is fundamentally a non-linear combination of $y'$ and $y''$ which have different dimensions, and constraining its value forces these quantities to vary together during reconstruction. Decoupling the first and second derivatives of the polynomials enables more intuitive scaling of the optimization parameters and, empirically, yields lower average fitting errors.

As in \cite{yair2025shapecal}, these relations provide $5N-7$ equations, whereas there are $5N-4$ unknowns. The remaining three equations are obtained by fixing one array endpoint at a predetermined position---$[x_0,y_0]=[0,0]$ in our case---and constraining the other endpoint to lie on the $x$-axis, namely $y_{N-1}=0$. In practice, the additional constraints resolve the problem's invariance to translation and rotation \cite{Dokmanic2015edm, ShapeCal, braaten2013selflex} common to other shape estimation methods, and provide the absolute reference position and antenna plane orientation needed for radiation pattern synthesis. The system of equations can be solved numerically using the algorithm presented in \cite{Levenberg1944lm}, which is readily available in standard scientific programming suites. Compared with previous segmentation-based shape estimation methods \cite{ODonovan1973strain_array} in which segment positions are determined iteratively one after the other, our formulation accounts for all measurements simultaneously, thereby enhancing robustness under measurement noise. This advantage is demonstrated in the statistical evaluation presented in Section \ref{sec:sampling}. Permitting the arc lengths in \eqref{arc} to be free design variables in the system enables to tailor the sensor placement to the desired application---for example, at expected folding points to improve accuracy---or to apply the method to rigid–flex arrays based on the known positions of folding joints. Another feature of the proposed method is its applicability to 2-D flexible surfaces. A grid of horizontal and vertical curvature sensors can resolve a grid of polynomials across the surface, serving as a basis for 2-D interpolation \cite{De_Boor2001splines} to reconstruct the array geometry. This approach avoids challenges faced by 2-D RF-coupling–based reconstruction methods \cite{Oren2026_2D}, which may suffer from inhomogeneous interference effects and require more complex, computation-intensive processing. Although not implemented here, this extension follows directly from the same framework and can be pursued in future work.
 
\section{Data sampling and sensor placement}
\label{sec:sampling}
To evaluate the effectiveness of shape reconstruction, we define the reconstruction error $Err_{av}$ as the average Euclidean distance between the coordinates of the sampled points and their estimated positions, normalized by the operating wavelength $\lambda$. For a narrowband wavefront propagating in the direction $\hat{r}$, with wavenumber $\vec{k}$, angular frequency $\omega$, and accumulated phase
    \begin{equation} 
    \label{eq:phase_prop}
    \varphi = (\omega t-\vec{k} \cdot \vec{r}),
    \end{equation}
this definition corresponds to a phase error in the transmitted signal, $\varphi_{err}=2\pi \cdot Err_{av}$, which is an important performance metric in wireless systems. Although phase error is often reported in rms terms, we use the average deviation to maintain consistency with \cite{ShapeCal,oren2021shapecal,yair2025shapecal,austin2022jom_flex}. In the absence of additional knowledge about the array’s structure, our analysis assumes homogeneous sampling along the array curve. One exception concerns the outermost sensors, which cannot be placed at the array endpoints because of their finite dimensions. Moreover, the force required to bend the array near its ends is significantly higher than elsewhere, resulting in limited curvature and reduced measurement sensitivity in those regions. Consequently, these sensors are nominally positioned at the midpoints of their respective segments. Thus, in the na\"ive case,
    \begin{equation}
    \label{eq:endpts_sample}
    s_i =
    \begin{cases}
    s_{\mathrm{total}}/N, & 1<i<N-1 \\
    s_{\mathrm{total}}/2N,  & i=1,N
    \end{cases}.
    \end{equation}
Next, we analyze the dependence of $Err_{av}$ on $N$ when reconstructing singly- and doubly-curved 1-D arrays from curvature measurements. These cases can later be extended, without loss of generality, to represent shapes with a larger number of bends. We do not consider arrays with predetermined sharp bends, which should use dedicated sensors placed at the bend locations, nor arrays with non-predetermined sharp bends, which may compromise signal integrity and mechanical reliability in flexible PCBs. Classical sampling analyses typically require either a known finite-dimensional parametrization of the admissible shape family or a tractable function class for which reconstruction error can be bounded analytically. In our case, however, the array shape is unknown and inferred indirectly from the sampled curvature under practical sensor-placement constraints. The purpose of the analysis here is to evaluate the sensor density required to achieve the desired performance in realistic flexible RF-PCB arrays rather than to derive a universal sampling theorem. Therefore, we chose to perform an empirical analysis in which we generated a large set of curves, applied our reconstruction procedure to them at different sampling rates, and inspected the quality of the estimation results. 
\begin{figure}[t]
\centering\includegraphics[clip, trim=0cm 0.2cm 0cm 0cm, width=\linewidth]{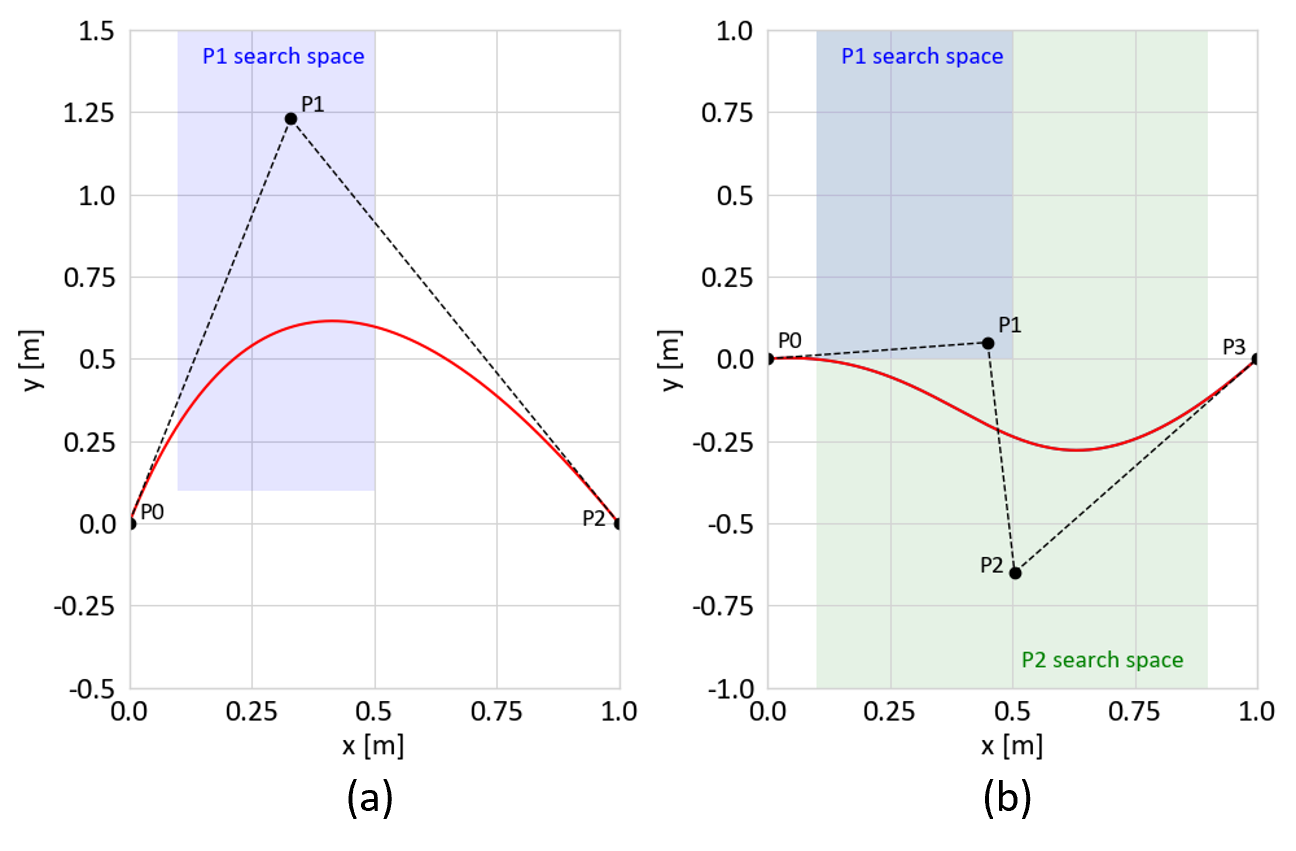}\caption{Control point sample space, and exemplary (a) second-order and (b) third-order B\'ezier curves used for performance analysis of the reconstruction algorithm.}
\label{fig:bezier_example}
\end{figure}

To generate the dataset, we restricted the admissible curves to those consistent with practical properties of existing flexible RF PCBs: a minimum radius of curvature of approximately 20 cm ($k_{max}\approx5$ m$^{-1}$) for meter-scale arrays \cite{matan2022npj}. Since our method is designed to evaluate shapes of functions $y(x)$ \cite{yair2025shapecal}, we further restricted the dataset to shapes with slopes $|y'|\leq10$, beyond which the cubic-polynomial fit used here may encounter difficulties. We generated a total of 11,000 1-D curves on a normalized shape space with their endpoints at (0, 0) and (1, 0) in the x--y plane. 1,000 curves were parametrized as second-order B\'ezier curves, and 10,000 curves were parametrized as third-order B\'ezier curves, with either a single or double bend. The use of third-order curves enabled us to add asymmetry to the shapes while avoiding overly sharp bends. Example curves and the search space used to generate them are illustrated in Fig. \ref{fig:bezier_example}. Since the reconstruction algorithm is invariant to mirroring about the y-axis and about the vertical centerline ($x=0.5$), the first control point $\mathrm{P1}$ was chosen from a search space within the range $0.1\leq x \leq 0.5$, $0.1 \leq y \leq 1$. For second-order B\'ezier curves, the vertical search space shown in Fig. \ref{fig:bezier_example}a was slightly extended to $0.1 \leq y \leq 1.5$ to enable the generation of additional valid curves. Once $\mathrm{P1}$ is chosen, the shape space is no longer symmetric with respect to the choice of $\mathrm{P2}$, which, for third-order B\'ezier curves, was chosen from the range $0.1\leq x \leq 0.9$, $-1 \leq y \leq 1$, as shown in Fig. \ref{fig:bezier_example}b.

\begin{figure}
\centering
\includegraphics[width=1\linewidth]{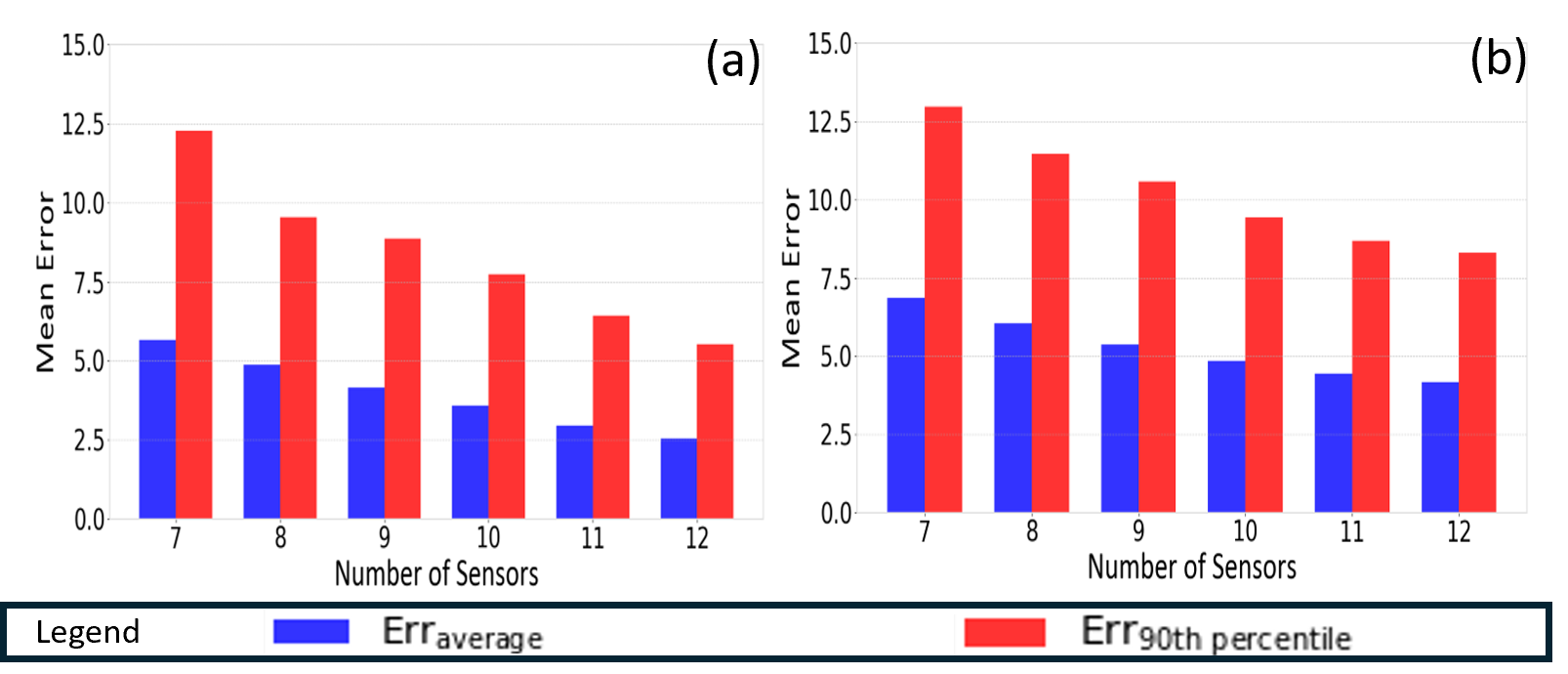}
\caption{Reconstruction error statistics for 2nd (left) and 3rd (right) order B\'ezier curves: mean error relative to $\lambda$ (blue) and mean-error-dispersion (red).}
\label{fig:sensor_statistics}
\end{figure}

We applied our reconstruction algorithm to all the generated curves for sampling densities $N$ = 7 to 12 sensors, and calculated $Err_{av}(N)$. Since $Err_{av}>0$ and most errors are small compared to their spread, the error distribution is non-Gaussian, and the standard deviation is not representative of the error statistics. We therefore inspected the reconstruction error below which 90\% of samples fall, which we term the maximum error dispersion (MED). Calculation results are summarized in Fig. \ref{fig:sensor_statistics}. For both 2nd- and 3rd-order B\'ezier curves, both the average and the MED of $Err_{av}$ decrease monotonically with the sampling density, as expected. Notably, for $N$=9, the weighted average of the errors in Figs. \ref{fig:sensor_statistics}a and \ref{fig:sensor_statistics}b is $Err_{av}\approx5\%$ and MED$\approx10\%$, yielding performance similar to or better than that of other published shape-reconstruction methods. As a result, we chose to demonstrate our algorithm using nine curvature sensors, dividing an eight-element array into eight analytically reconstructed segments.

\begin{figure}
\centering
\includegraphics[width=1\linewidth]{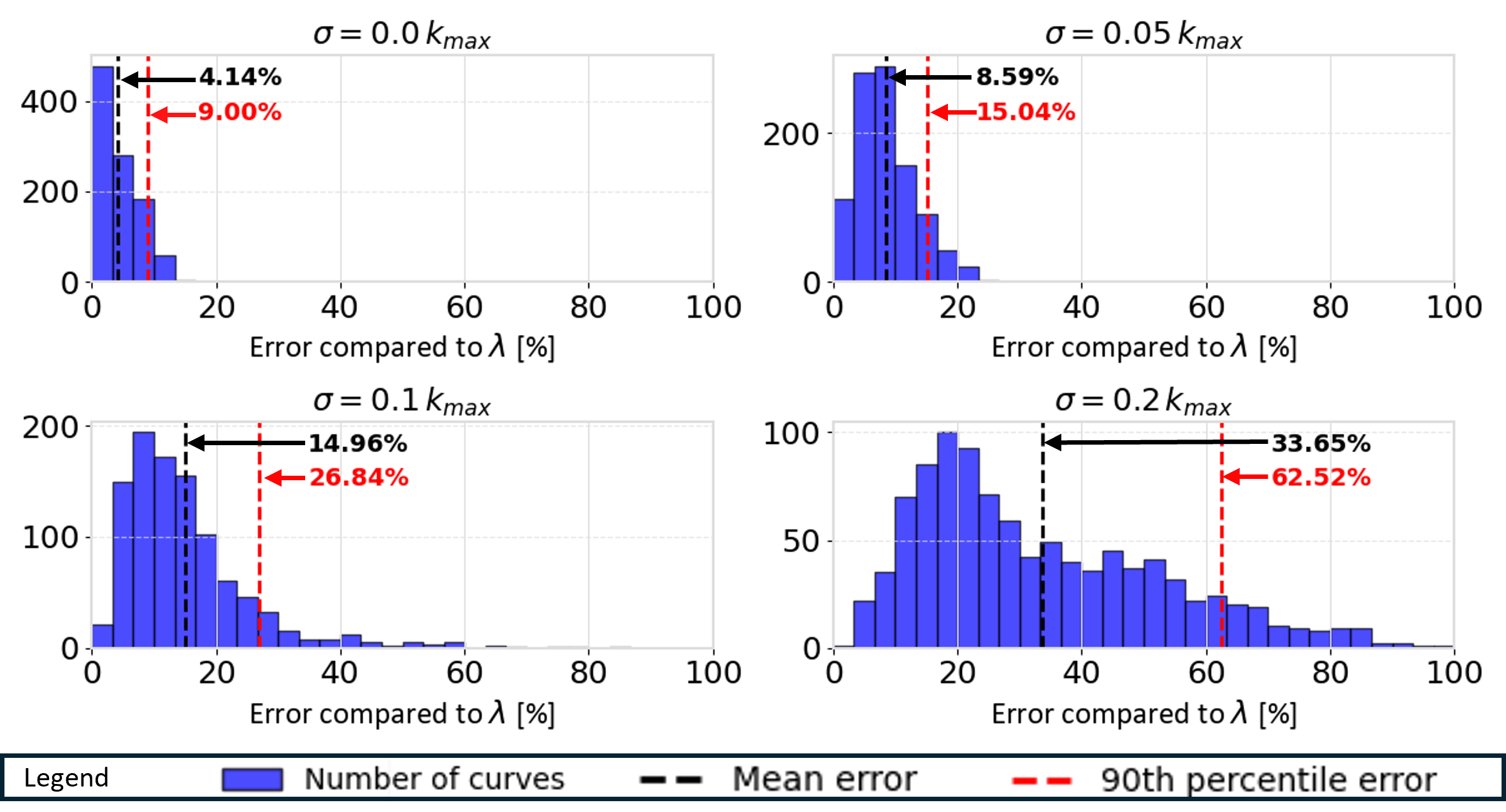}
\caption{Reconstruction error histogram for 2nd-order B\'ezier curves using 9 sensors under Gaussian noise $\sigma\cdot k_\text{max}={[0,0.05,0.1,0.2]\cdot5}$. Black dashed line: Mean error; Red line: 90th percentile error.}
\label{fig:third_ord_bezz_noise}
\end{figure}
\begin{figure}
\centering
\includegraphics[width=1\linewidth]{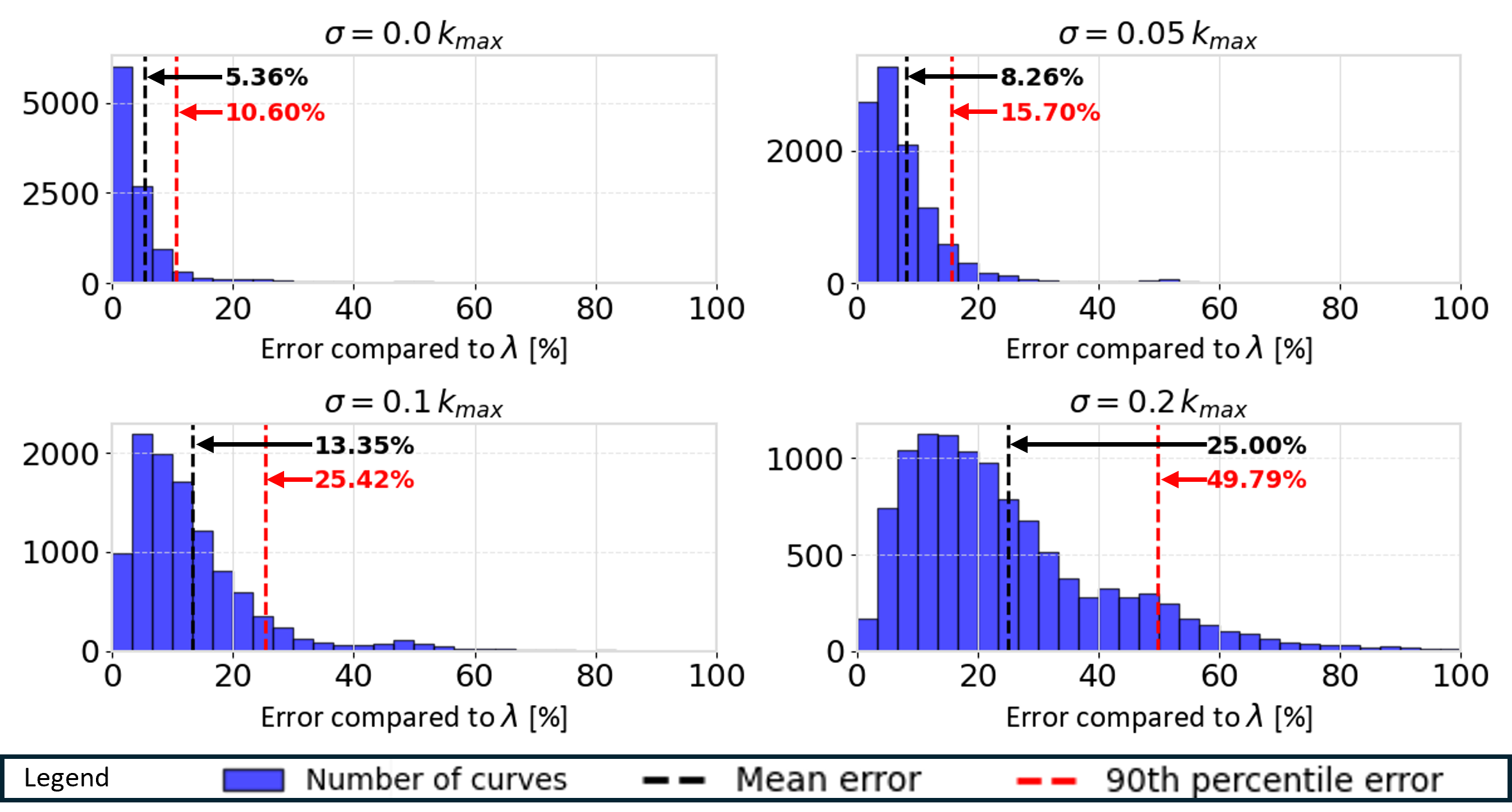}
\caption{Reconstruction error histogram for 3rd-order B\'ezier curves using 9 sensors under Gaussian noise $\sigma\cdot k_\text{max}={[0,0.05,0.1,0.2]\cdot5}$. Black dashed line: Mean error; Red line: 90th percentile error.}
\label{fig:fourth_ord_bezz_noise}
\end{figure}

To assess noise robustness, we added random noise to each curvature measurement in the dataset before running the reconstruction algorithm such that 
\begin{equation}
\label{eq:noise_addition}
\vec{k}=\vec{k}_{\text{meas}}+k_{\text{max}}\cdot \vec{\sigma}_\text{n},
\end{equation}
where $\vec{k}_{\text{meas}}$ are the ideal noiseless curvature samples in each curve, $\vec{\sigma}_\text{n} \sim \mathcal{N}(0, \sigma^2)$, and $k_\text{max}$ is a scale factor normalizing the error to the full curvature scale. We then reconstructed each curve in the dataset using the noisy curvature samples $\vec{k}$ and produced histograms of reconstruction errors over the curves in our $N=9$ dataset for $\sigma=[0, 0.05, 0.1, 0.2]$. Importantly, $\sigma$=0.05 represents a moderate measurement error, with $\sim$70\% of the added noise falling within $\pm$5\% and an additional $\sim$30\% of the added noise falling within $\pm$5\% to $\pm$15\% of the full scale of the expected curvatures. Likewise, $\sigma$=0.2 is an extreme case, in which many of the errors fall within a range as wide as the maximum span of expected curvature measurements. Figs. \ref{fig:third_ord_bezz_noise} and \ref{fig:fourth_ord_bezz_noise} illustrate the reconstruction error histograms for the 2nd- and 3rd-order B\'ezier-curve sets, where large measurement errors of $\sigma$=0.1 result in reduced reconstruction accuracy of $Err_{av}\approx15\%$. However, the reconstruction algorithm is expected to maintain reasonable performance under moderate measurement error of $\sigma$=0.05, with $Err_{av}\approx8\%$ and MED$\approx15\%$. As demonstrated later in this work, such measurement accuracy is usually achievable.

\section{Flexible antenna array design} 
\begin{figure}
    \centering
    \includegraphics[width=1\linewidth]{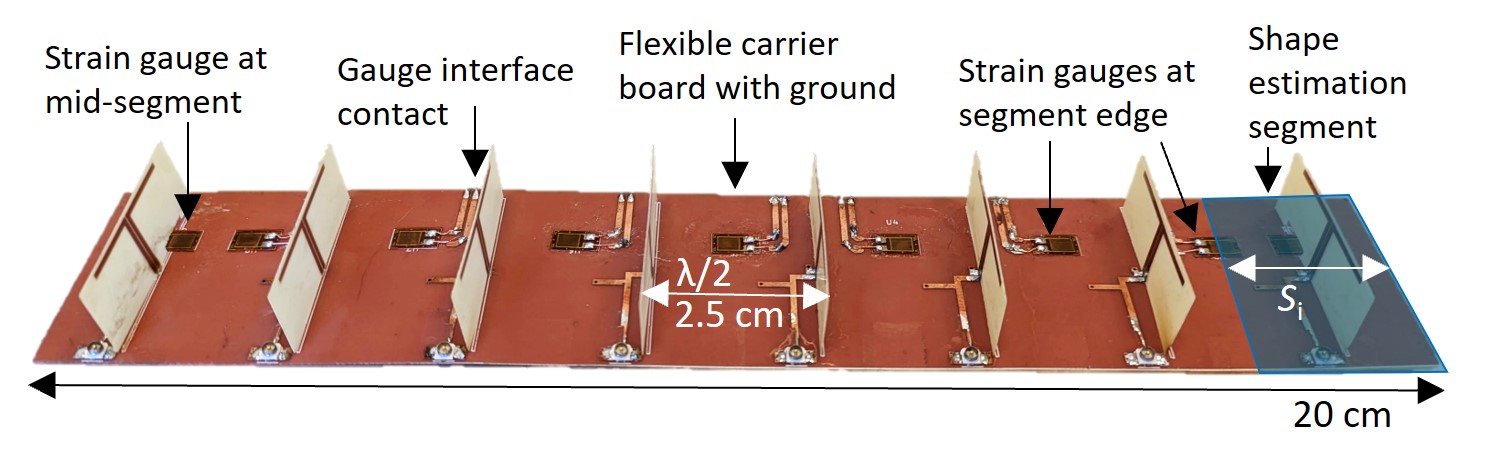}
    \caption{Final assembly of the flexible phased array with sensors connected to their designed traces and the radiator elements connected to the interface UMMC connectors.}
    \label{fig:flex_array_build}
\end{figure}

To demonstrate the proposed method, we designed a flexible antenna array with integrated curvature-sampling capability, as shown in Fig. \ref{fig:flex_array_build}. The array comprises eight radiating elements spaced $\lambda/2$ apart, and operates at frequencies around 6 GHz, which are widely used in modern 5G, Wi-Fi, and wireless-video applications. This element count enables the formation of a sufficiently narrow main lobe to clearly illustrate beam focusing and steering. The array was fabricated on a 0.175-mm-thick FR-4 laminate with two metal layers, allowing significant surface deformation---with local radii of curvature below 3.5 cm---without causing permanent damage to the circuit board. Following the sampling considerations of Section \ref{sec:sampling}, nine strain gauges were integrated along the array to serve as curvature sensors. Importantly, the flexible antenna itself does not include phase- or delay-tuning circuitry; instead, it relies on preconditioned signals fed to the individual radiators to realize the phased-array functionality. This intentional design choice reduces component count and decouples uncertainties in the RF signal path from the core shape-reconstruction functionality that this work aims to demonstrate.

\subsection{Radiator design}
\begin{figure}
    \centering
    \includegraphics[width=1\linewidth]{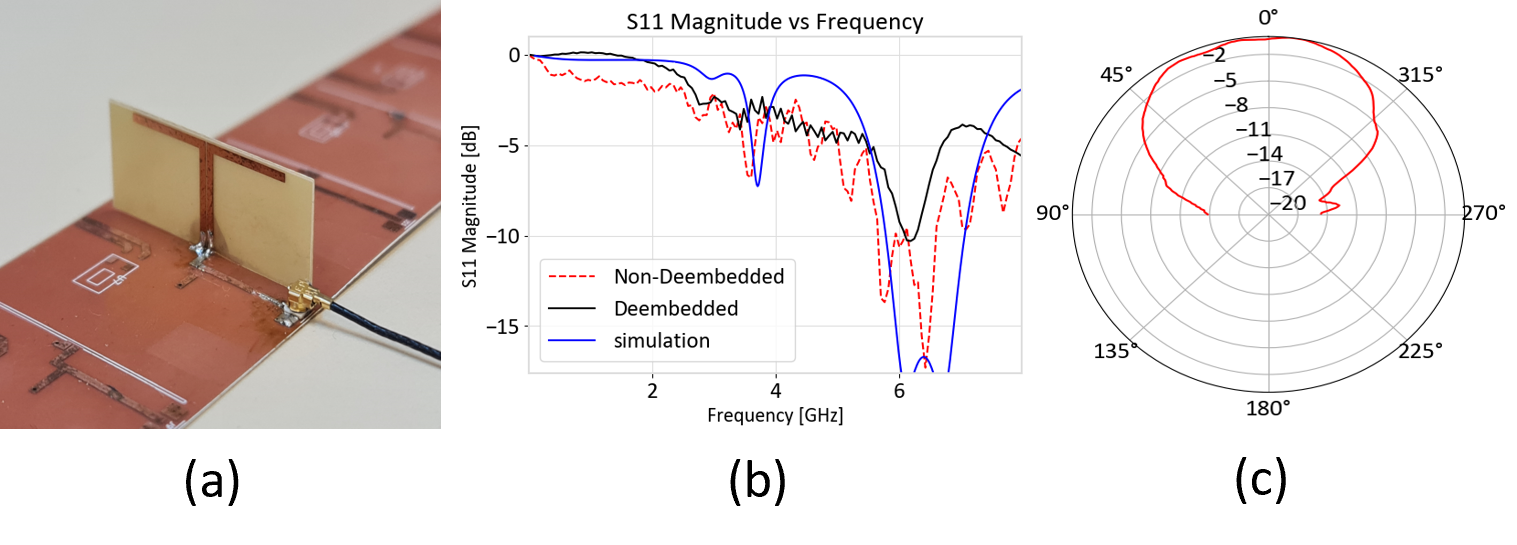}
    \caption{Single radiator performance. (a) Thin FR-4 antenna, 22.9 mm $\times$ 11 mm, (b) input matching with and without a 15-cm coaxial cable connected at the input, and (c) the radiation pattern of a stand-alone single radiator.}
    \label{fig:single_antenna_design}
\end{figure}

The radiators are $\lambda/2$ dipoles, printed on a thin single-sided, 0.175 mm FR-4 laminate, and soldered perpendicular to the flexible array's ground plane (Fig. \ref{fig:single_antenna_design}a). The vertical $\lambda/4$ coupled line connecting the radiator to the array board matches its input impedance to standard 50$\Omega$. Each radiator is fed from a separate U.FL connector to support individual phase steering. Matching is characterized in Fig. \ref{fig:single_antenna_design}b, which compares simulation, a non-de-embedded measurement using a 15 cm SMA to U.FL adapter cable, and a de-embedded measurement using short-open-load calibration to the board plane after the connector. The radiation pattern of the individual radiator, shown in Fig. \ref{fig:single_antenna_design}c, was measured in an anechoic chamber, exhibiting a half-power beamwidth of $\sim$90$^{\circ}$. The use of dipole antennas serves several purposes: the thin structure allows free 1-D shape deformation, and the broad beamwidth assists in maintaining radiation performance even when the dipole is not perpendicular to the intended steering direction. Using these antennas also maintains compatibility with \cite{oren2021shapecal,ShapeCal,yair2025shapecal}, for a close comparison of performance between the proposed methods.
\subsection{Strain gauges as curvature sensors}
\begin{figure}[b]
    \centering
    \includegraphics[clip, trim=0cm 0cm 0.3cm 0cm, width=\linewidth]{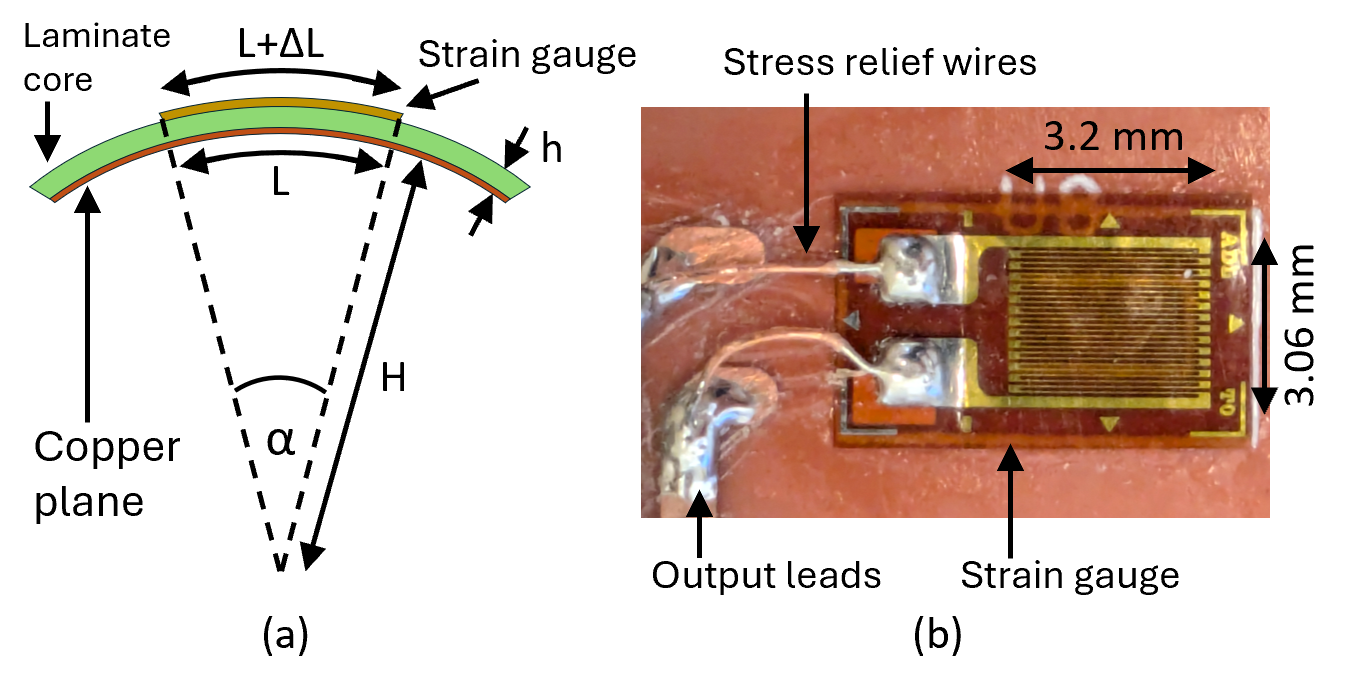}
    \caption{(a) Effect of bending a curvature sensor mounted onto a PCB, and (b) a strain gauge mounted on to the flexible antenna array.}
    \label{fig:sensor_on_curve}
\end{figure}

In this work, we measure discrete, local curvature samples along the array using surface-mounted strain gauges. We assume that bending the array produces mostly lateral strain, and that when bent, the copper ground layer strains significantly less than the top, mostly copper-free layer. The latter assumption is justified by the difference in Young’s modulus between copper, $\sim$130 GPa \cite{Freund2010materials}, and FR-4, $\sim$20-30 GPa \cite{fr4_youngs}. Under these assumptions, the curvature is approximately proportional to the strain of the top surface of the circuit board. Fig. \ref{fig:sensor_on_curve}a illustrates how local curvature and strain are related: a strain gauge with a length $L$ attached to the top of a flexible board with a thickness $h$ is strained to a total length of $L+\Delta L$ when the board is bent with a local curvature $H$.
The angle that the board occupies can be calculated from the length of the ground plane,
\begin{equation}
\label{l_length}
    L = \alpha H,
\end{equation}
but also from the elongated top plane,
\begin{equation}
\label{ldl_length}
    L+\Delta L = \alpha(H+h).
\end{equation}
Dividing \eqref{ldl_length} by \eqref{l_length} and substituting the curvature $k=1/H$ yields the linear relation
\begin{equation} 
\label{eq:strain_vs_k}
    \frac{\Delta L}{L} =  h \cdot k.
\end{equation}
In this work, we utilize commercially available, 350$\Omega$, BF350-3AA gauges to measure strain in the required sample points. The sensing body dimensions are 3.2$\times$3.06 mm--significantly smaller than the reconstructed segment size of 25 mm, enabling representative local curvature measurements. These sensors exhibit a strain limit of 2\%, which, following \eqref{eq:strain_vs_k}, allows us to measure a minimum curvature radius of 1 cm in 0.2 mm-thick circuit boards without permanent mechanical damage.
As shown in Fig. \ref{fig:sensor_on_curve}b, the sensors were bonded to exposed glass-epoxy areas on the top surface of the flexible circuit board using cyanoacrylate adhesive to ensure a long-term bond, while adding negligible adhesive-layer thickness. After bonding, the sensor pads were interfaced with the traces used for measurement contacts by soldering thin, loose, copper wires between them---a standard practice in strain gauge installation aimed at relieving excess stress caused by movement of the rigid interface traces.

\subsection{Curvature measurement and sensor calibration}
\begin{figure}
    \centering
    \includegraphics[width=1\linewidth]{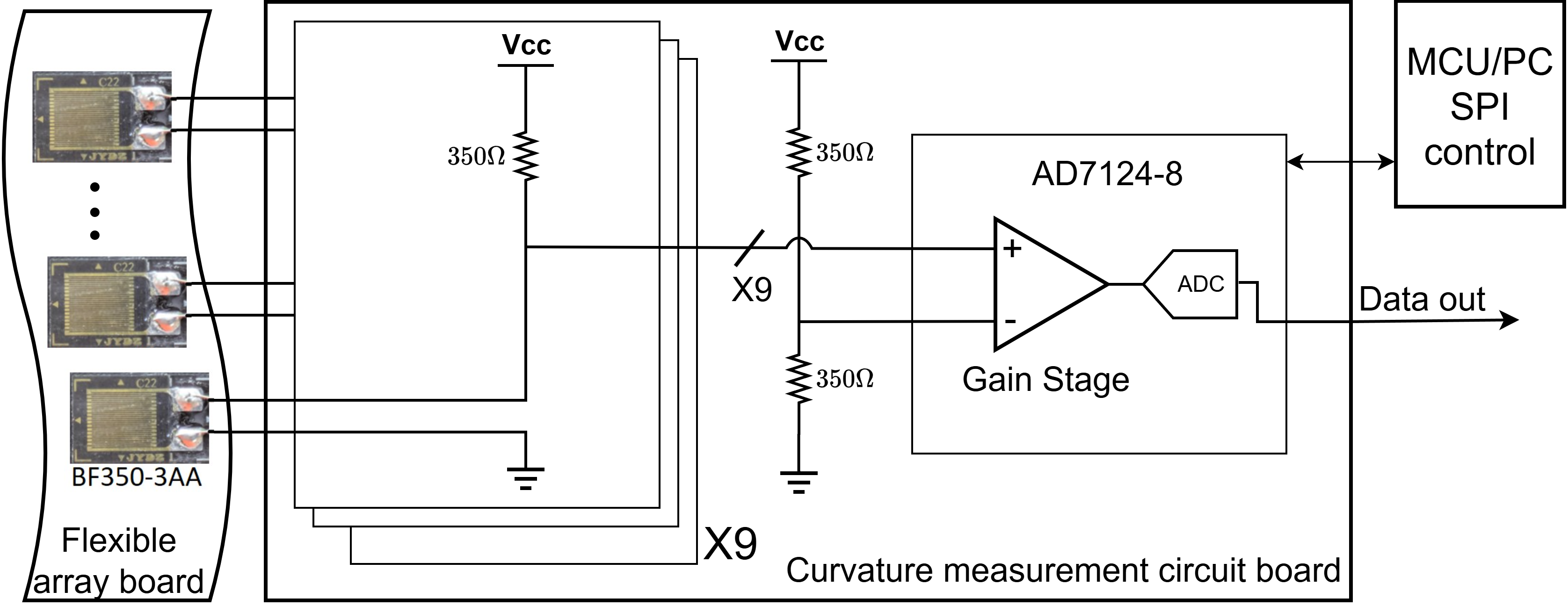}
    \caption{Multi-input strain gauge measurement circuit based on AD7124-8 ADC.}
    \label{fig:Characterization_circuit}
\end{figure}
The strain gauges were measured using a multichannel analog-to-digital converter (ADC), AD7124-8, implemented on a separate circuit board, as shown in Fig. \ref{fig:Characterization_circuit}. This ADC enables consecutive measurement of all nine sensors in a single command, reducing the errors resulting from manual switching between measured sensors observed in \cite{yair2025shapecal}. In the current implementation, nine 350 $\Omega$ resistors are placed in proximity to nine of the ADC inputs, forming a voltage divider with each of the strain gauges when connected to the leads from the flexible antenna board. A tenth, additional input of the ADC is connected to a permanent resistive divider formed by two additional 350 $\Omega$ resistors, serving as a voltage reference. The ADC is then configured to perform a sequence of pseudo-differential measurements in which each strain gauge, in turn, forms a Wheatstone bridge with the reference branch. Since the sensors have a gauge factor of $\sim$2, the maximum permissible change in resistance is $\Delta R/R=\pm$4\%, which, in a Wheatstone bridge connected to a supply $V_{DD}$ with a normalized output voltage difference of
\begin{equation}
\label{eq:wheatstone_voltage_out}
    \frac{\Delta V_{out}}{V_{DD}} = \frac{R+\Delta R}{2R+ \Delta R}-\frac{1}{2} \approx \frac{\Delta R}{4R}
\end{equation}
translates to a maximum output voltage difference of 1\%. The expected differential voltage can be expressed as a function of curvature by replacing $\Delta R/R$ with the strain multiplied by the gauge factor, and substituting \eqref{eq:strain_vs_k} into \eqref{eq:wheatstone_voltage_out},
\begin{equation}
\label{eq:voltage_vs_k}
    \Delta V_{out} =V_{DD}\frac{\Delta R}{4R}\approx\frac{V_{DD}}{2}h\cdot k.
\end{equation}

\begin{figure}
    \centering
    \includegraphics[width=1\linewidth]{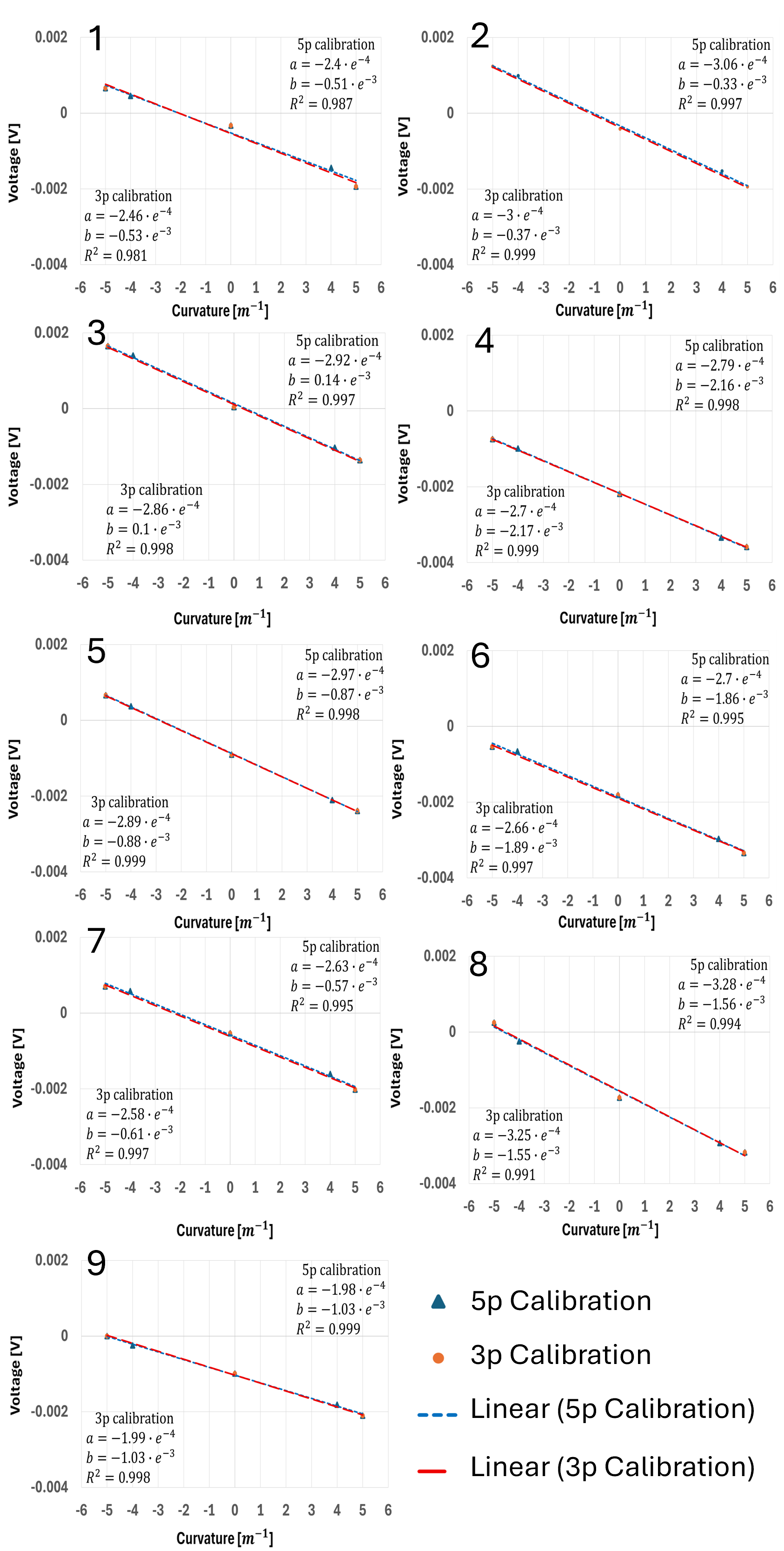}
    \caption{Calibration of each curvature sensor using shapes with known curvature values $k$ for three- and five-point linear response.}
    \label{fig:sensors_calibration}
\end{figure}
The small voltage differences can be sampled with sufficient resolution by utilizing the ADC's built-in pre-amplifiers and 24-bit conversion accuracy.
While the strain gauges show good linearity and temperature stability, small differences in responsiveness and nominal resistance still exist between sensors, which need to be accounted for before use. To do so, we measured the response of each gauge against several known curvatures. We produced five circular-arc shapes with constant curvatures $k=[\pm5.2, \pm4, 0]$ m\textsuperscript{-1} and a total arc length of 20 cm, conformed the array to each of them, measured all nine sensor output voltages within one automated acquisition sequence, and used linear regression to obtain the output voltage as a function of curvature. We also performed a simpler, faster calibration based on only three curvatures, where $k=[\pm5.2,0]$ m\textsuperscript{-1}. The ADC internal gain was set to $\times$128, and the measurement reference voltage was 3.3 V. The total board thickness was assumed to be 0.2 mm, considering the FR-4 core thickness and a post-production 1-oz copper ground layer, resulting in an expected slope of $0.5V_{DD}\cdot h=0.33\cdot10^{-3}$ V/m\textsuperscript{-1}. The raw measurements were divided by the ADC gain to compare the slope to the expected value from \eqref{eq:voltage_vs_k}. Results for each of the nine sensors are shown in Fig. \ref{fig:sensors_calibration}. The average measured slope over all internal sensors was $\sim 0.29\cdot10^{-3}$ V/m\textsuperscript{-1}, confirming our physical assumptions\footnote{Small differences are expected due to unaccounted non-idealities and imperfect sensor installation, e.g., over silk-screen.}, while the outermost sensors showed reduced sensitivity, as expected. All the gauges demonstrated linear behavior, with a worst-case $R^2=0.981$. On average, three-point calibration yielded a 1.4\% error in the slope and 2\% in the y-intercept compared to using the full dataset. This modest loss in accuracy might be acceptable in time- or labor-sensitive scenarios that prohibit executing the full five-point calibration.

\section{Experimental setup}
\subsection{Flexible array driver circuitry}
\begin{figure}[b]
    \centering
    \includegraphics[width=1\linewidth]{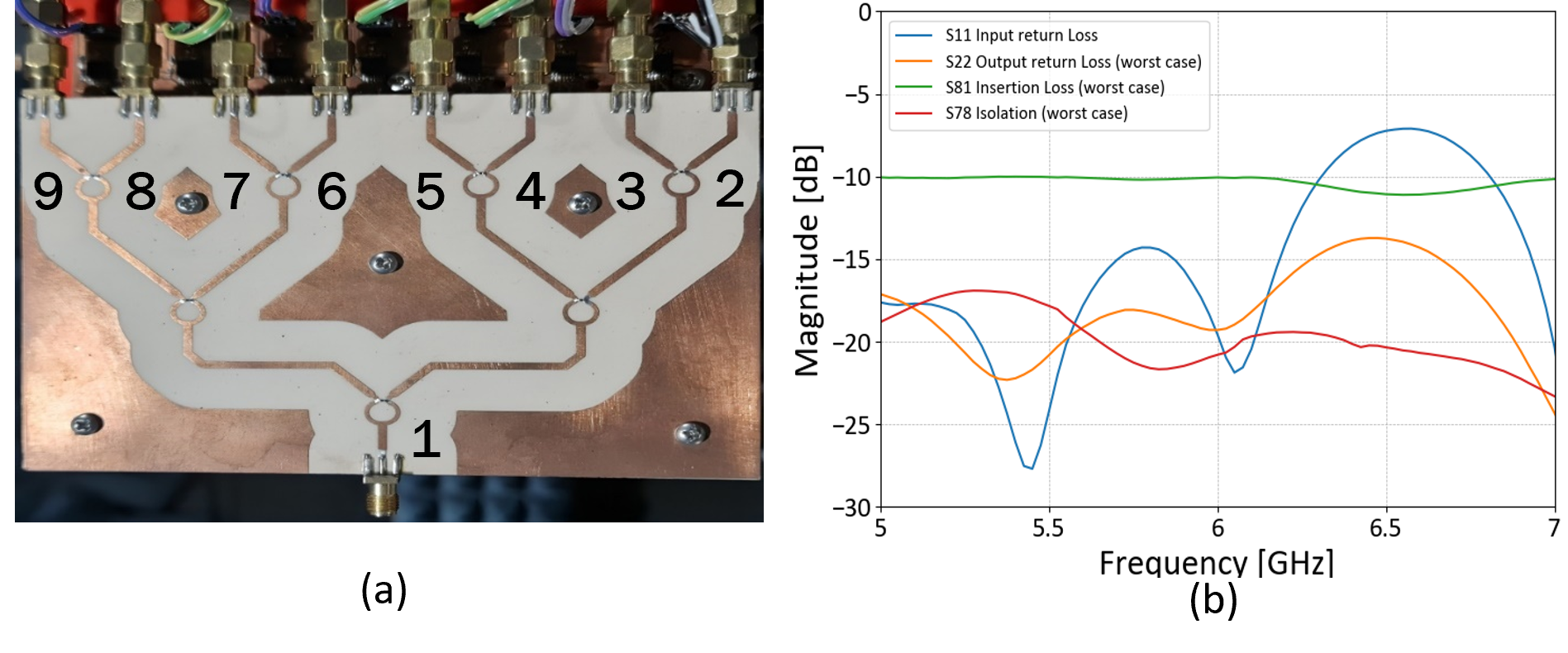}
    \caption{(a) The used three-stage Wilkinson splitter and (b) its measured performance (for worst-case channel and isolation)}
    \label{fig:Wilkinson splitter & meas}
\end{figure}
\begin{figure}[t]
    \centering
    \includegraphics[clip, trim=0cm 0.2cm 0cm 0cm, width=\linewidth]{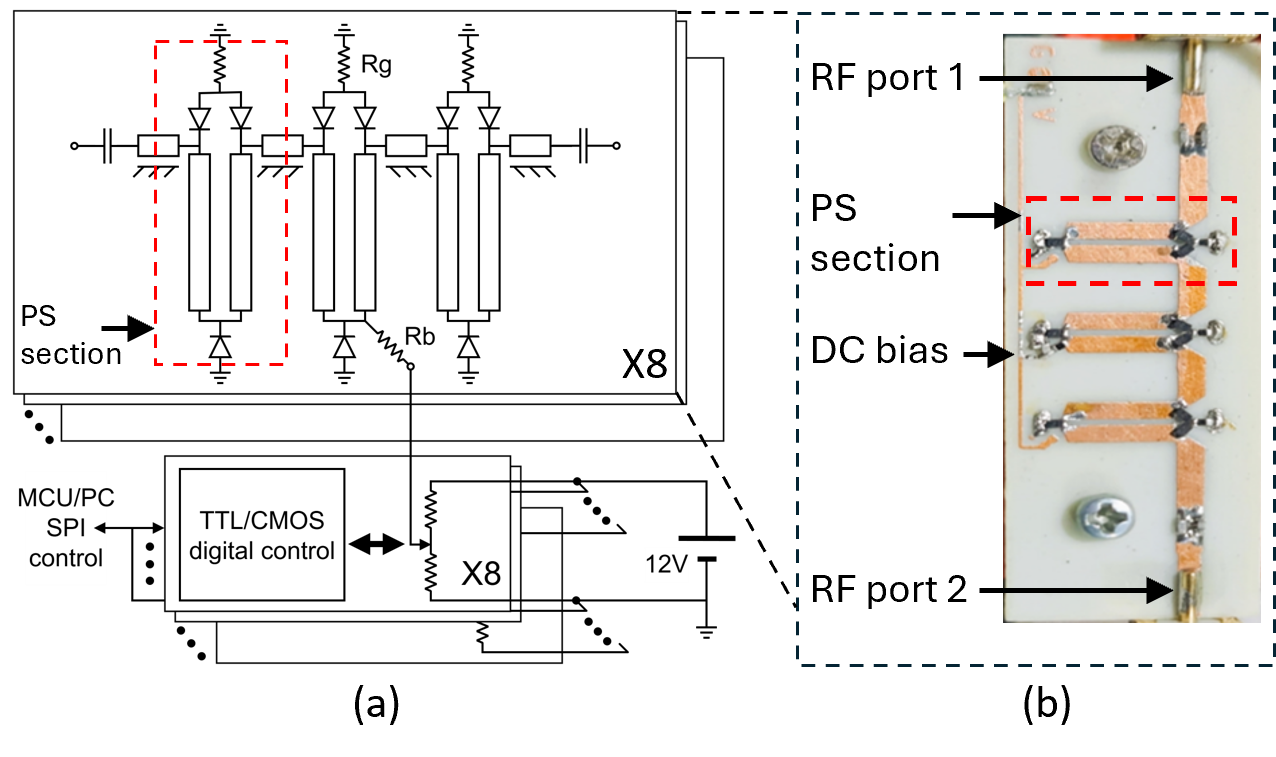}
    \caption{(a) Simplified schematic of the phase-shifter path and digitally-controlled bias circuitry. (b) Fabricated phase-shifter.}
    \label{fig:Phase_shifter_scheme&photo}
\end{figure}

\begin{figure}[b]
    \centering
    \includegraphics[clip, clip, trim=0cm 0.2cm 0cm 0cm,width=1\linewidth]{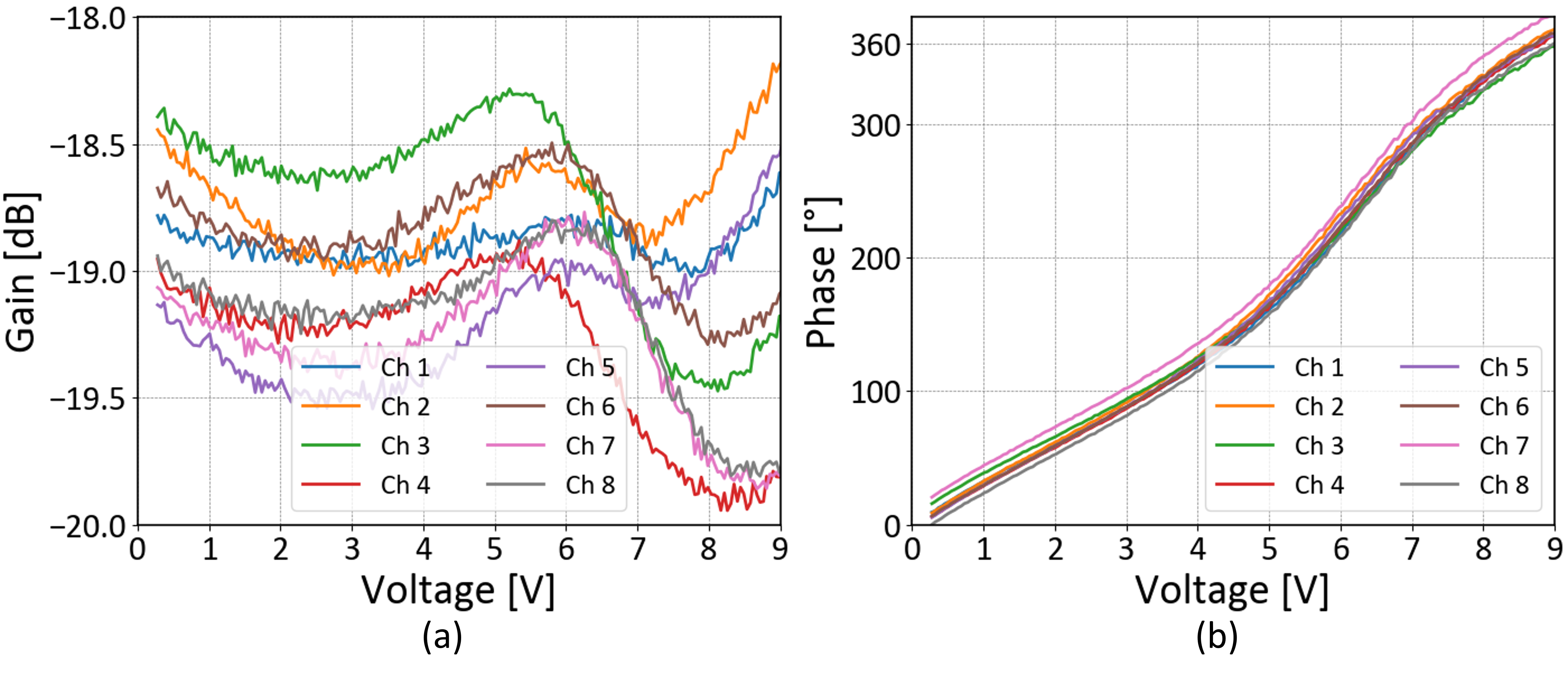}
    \caption{ Post-calibration (a) amplitude and (b) phase of the fabricated phase shifters as function of the applied control voltage}
    \label{fig:PS-characterization_w_splitter}
\end{figure}

The array driver circuitry implements the signal conditioning required to demonstrate phased array operation, including a 1:8 signal splitter and individual phase shifters for each of the flexible antenna's radiators.  While full control of beamforming parameters requires additional amplitude control for each channel, it was shown in \cite{braaten2013selflex,yair2025shapecal} that for flexible arrays with sufficiently isotropic radiators separated by $\sim \lambda/2$, acceptable steering of the main lobe can be achieved by controlling only the channel phases. We therefore chose not to implement amplitude control to minimize uncertainties in the signal path, and to simplify the control interface and required calibration. The splitter, illustrated in Fig. \ref{fig:Wilkinson splitter & meas}a, is a three-stage Wilkinson divider, implemented on a 30 mil RO4350 laminate. As shown in Fig. \ref{fig:Wilkinson splitter & meas}b, this implementation provides a broadband impedance match up to 6.3 GHz, with a measured insertion loss of $\sim$1 dB for each channel, and inter-channel amplitude variation $\leq$0.1 dB, including the connectors. The worst-case in-band channel isolation, between neighboring channels (e.g., $S_{78}$), is $\leq$-20 dB, preventing mutual effects between phase shifters.
The splitter outputs drive the phase shifters, which in turn feed the array radiators. The phase shifters, illustrated in Fig. \ref{fig:Phase_shifter_scheme&photo}, rely on the same all-pass topology presented in \cite{khoder2014ps}. However, instead of using two different all-pass sections tuned to different frequencies, they utilize three sections tuned to the same frequency of $\sim$6.1 GHz. This approach provides better in-band insertion-loss variation at the expense of a reduced bandwidth. The implemented all-pass circuit also replaces the originally-reported varactors, MA46H120, with SMV2019 diodes, substantially reducing the cost of components. As shown in Fig. \ref{fig:Phase_shifter_scheme&photo}, the control voltage is decoupled from the phase shifters with a large series resistance \cite{khoder2014ps}, enabling it to be set programmatically from a PC or an MCU interface using a digitally controlled potentiometer. In this work, we used eight potentiometers with an 8-bit resolution, dividing a supply voltage of 12 V to set the phase shifters' control voltage. We characterized the eight phase shifters at our intended operating frequency of 6 GHz. For each phase shifter we swept the control voltage and recorded the relative phase shift and loss using a network analyzer, producing a digital state table. Fig. \ref{fig:PS-characterization_w_splitter} shows the resulting loss and relative phase shifts of all channels driving the flexible antenna, including the splitter, cables, connector, and phase shifters. Each channel provides 360$^{\circ}$ phase shift with an effective resolution better than 7-bit, with a gain variation $<$1 dB between the eight channels over all phase states.
In a separate, standalone characterization over an extended range of 5.6-6.8 GHz, we measured a nominal phase shifter gain of -6.7$\pm$0.2 dB with a maximum variation of $\pm$1.1 dB over all phase states. The minimum phase shift remained $>$360$^{\circ}$ and monotonic with the control voltage, and once characterized post-production, exhibited worst-case phase errors of $\sim4^{\circ}$ across 160 MHz-wide frequency bands in this frequency range. Even though the measured loss is larger than in some other implementations \cite{khoder2014ps,Padilla2016reconf_ps,MAPS010165}, the ease of design and tuning, together with the wide frequency range of operation, make these phase shifters attractive for cost-sensitive, calibration-tolerant research and prototyping arrays targeting sub-7 GHz applications, including Wi-Fi 6E, Wi-Fi 7, and 5G FR1 studies. 

\subsection{Beam Correction}
\begin{figure}[t]
    \centering
    \includegraphics[clip, trim=0cm 0.5cm 0cm 0.8cm,width=0.75\linewidth]{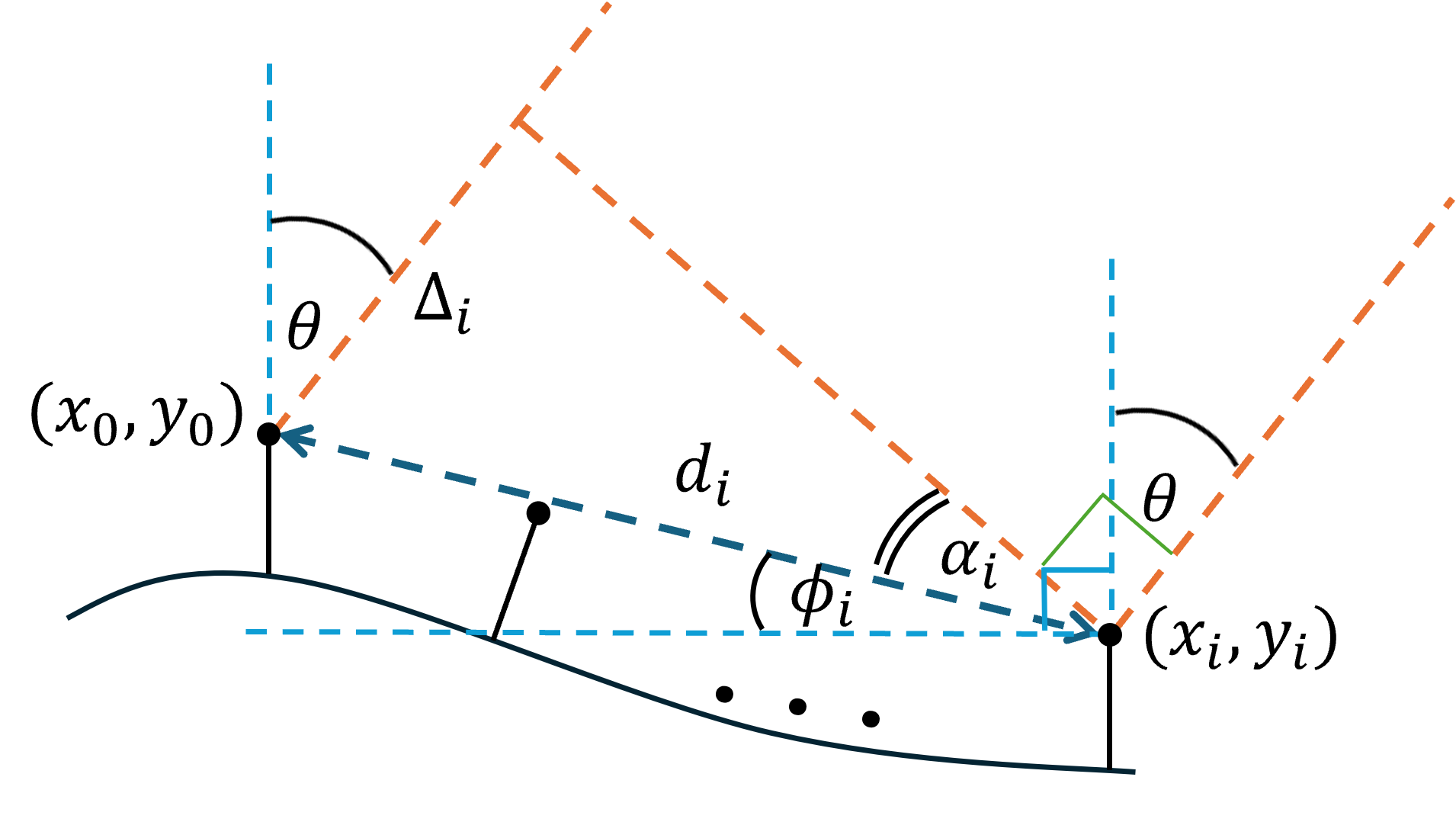}
    \caption{Illustration of phased-array element locations and their corresponding phase corrections.} 
    \label{fig:Phased array correction}
\end{figure}
\label{beam_correction}
To demonstrate how shape reconstruction can be used to mitigate deformation effects in flexible phased arrays, we first calculated the phase adjustment at each element required to compensate for positional variation due to bend. This can be done similarly to the calculation of the beam steering angle of a planar array \cite{Balanis2016antenna}. As illustrated in Fig. \ref{fig:Phased array correction}, $\Delta_i$ is the path delay between element $i=1,2,\ldots,M-1$ in the array and a reference element at $(x_0,y_0)$, for a desired steering angle $\theta$ from broadside. Assuming all element positions are known after reconstruction, their distance $d_i$ from the reference can be calculated from
\begin{equation} 
    \label{distance}
    d_i = \sqrt{(x_i-x_0)^2 + (y_i-y_0)^2}.
\end{equation}
Correspondingly, the angle $\phi_i$ of the line segment of length $d_i$ relative to the reference plane is
\begin{equation} 
    \label{phi_angle}
    \phi_i = \tan^{-1}\left(\frac{y_i-y_0}{x_i-x_0}\right),
\end{equation}
and because $\alpha_i=(\theta-\phi_i)$,
\begin{equation} 
    \label{length}
    \Delta_i = d_i\cdot \sin(\theta-\phi_i).
    \end{equation}
For a narrowband signal, the phase delay $\varphi_i$ that requires compensation and the path delay $\Delta_i$ are related by $\varphi_i/2\pi=\Delta_i/\lambda$, therefore
\begin{equation}
    \label{eq:phase_correct}
    \varphi_i=\frac{2\pi}{\lambda} d_i \cdot \sin(\theta-\phi_i).
\end{equation}
Notably, this procedure compensates for deformation effects in flexible arrays by adding a phase shift to the radiated signal, which, as discussed, is valid when the single element patterns are approximately isotropic in the steering direction and the interelement spacing is $\sim\lambda/2$.

\begin{figure}[b]
    \centering
    \includegraphics[width=1\linewidth]{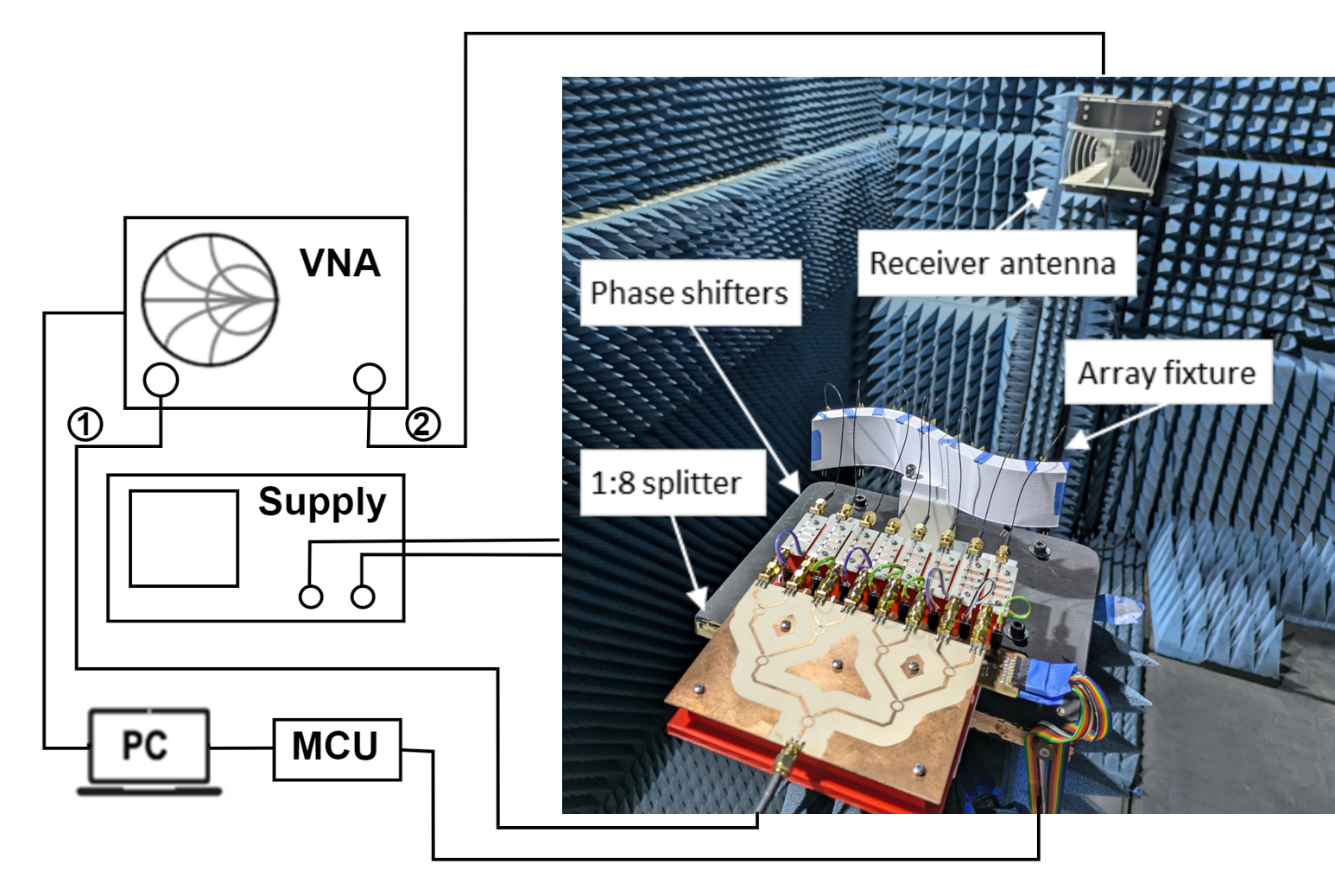}
    \caption{Experimental setup of the measurement apparatus in an anechoic chamber}
    \label{fig:experiment_setup}
\end{figure}

\section{Measurement results}
Our proposed reconstruction algorithm was evaluated in two different experiments. The first experiment is intended to verify our analysis in Section \ref{sec:sampling} against real measurements, while the second experiment demonstrates the method's applicability in reconstructing naturally occurring shapes that are not formed by a lower-dimension set of equations.

\begin{figure*}
    \centering
    \includegraphics[width=0.935\linewidth]{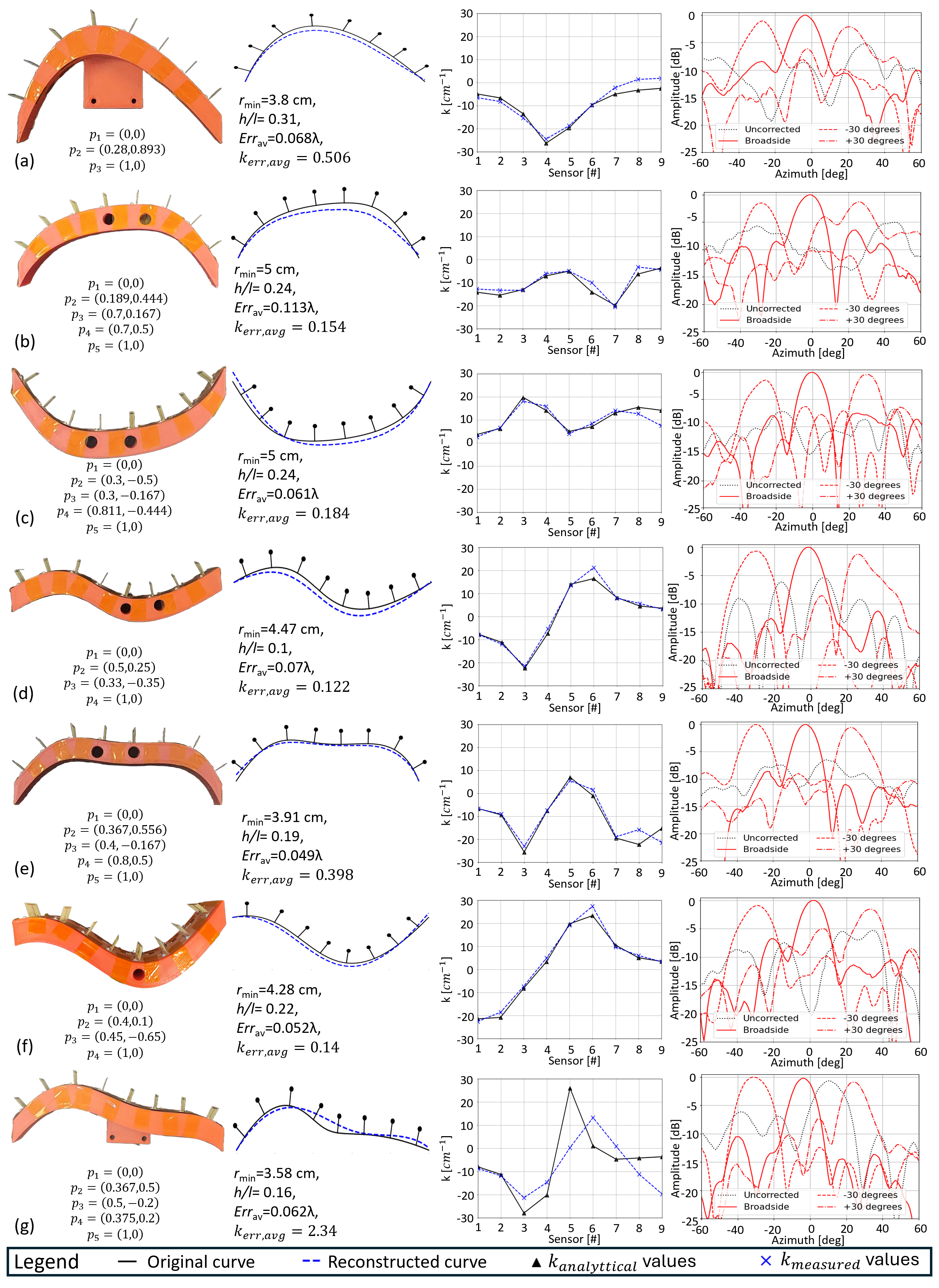}
    \caption{(Left to right) 3D printed B\'ezier shapes with phased array mounted upon with control points, shape reconstruction using curvature measurements compared to analytical curve with reconstruction error, $k_{measured}$ compared to $k_{analytical}$ and beam reconstruction and steering of each shape.}
    \label{fig:extruded_shapes_results}
\end{figure*}

\subsection{Flexible array conformed to curved surfaces}
The first experiment expands on the measurements reported in \cite{yair2025shapecal} by reconstructing the shapes of additional distinct, known 1-D curves. We produced several antenna fixtures with predefined 1-D shapes based on B\'ezier curves up to the fourth order, extruded and 3D-printed to form solids onto which the flexible array could conform. Similar to our previous work, we used each shape's minimum radius of curvature and the ratio between its height $h$ and its horizontal length $l$ as measures of the amount of bend. As illustrated in Fig. \ref{fig:experiment_setup}, the antenna was conformed to the fixtures, one at a time, and the curvature-sensor outputs were connected to the multi-channel ADC inputs. Initially, we measured all the curvatures along a given shape throughout a single acquisition sequence. We reconstructed the shape using our algorithm, obtained the radiator positions, and calculated the phase adjustment required at each radiator to steer the beam to broadside and to $\pm$30$^{\circ}$ relative to the main axis. We then connected the antenna, still conformed to the tested shape, to the phase-shifter outputs, using eight identical 15-cm U.FL cables with similar gain and phase delay. The phased array module comprising the splitter, the phase shifters, and the flexible array was measured using a rotating plate inside an anechoic chamber. To set a baseline, we set the phases of all the radiators to 0$^\circ$ and measured the array's 'uncorrected' 1-D radiation pattern. We then applied to each radiator the phase settings required to steer the beam in the desired directions for its reconstructed shape, measured the radiation patterns of the bent array, and compared them to the uncorrected state. Measurement results for all the shapes under test are shown in Fig. \ref{fig:extruded_shapes_results}. The second column from the left compares the analytical curves used to produce the antenna fixtures to the reconstructed curves based on curvature sampling. The tested shapes are more strongly curved than those in most other flexible phased array implementations \cite{ShapeCal,Poolakkal2025flex_nature_comm,Oren2026_2D}, with minimum curvature radii below $R_{\min}=3.6$ cm, and maximum $h/l$ ratios of 0.31. The average $Err_{av}$ over all twelve tested curves, including those in this work and in \cite{yair2025shapecal}, is 6.1\%, of which only one of the reconstructions exhibits $Err_{av}>10\%$. These values are in line with the average error of 5\% and MED of 10\% predicted in Section \ref{sec:sampling}. The differences are likely due to the small sample size compared to simulations, as well as additional unaccounted error factors such as 3-D production variations relative to the analytical profile, imperfect sensor installation, and unintentional deformations perpendicular to the curve axis. The third column from the left in Fig. \ref{fig:extruded_shapes_results} shows the raw sensor-curvature readings compared to those calculated from the analytical curves.
The effect of measurement errors on reconstruction was compared with the prediction of Section \ref{sec:sampling}. After re-normalizing the measured curvatures to the 1 m x-axis length assumed in the analytical framework, the average of the standard deviations calculated for each set of measurement errors in Figs. \ref{fig:extruded_shapes_results}a-\ref{fig:extruded_shapes_results}f is 7.5\%, which predicts slightly worse reconstruction errors than those measured in practice: $Err_{av}=8.3\%-13.3\%$. This result is likely because, unlike the AWGN measurement-noise assumption made in \eqref{eq:noise_addition}, practical errors are not uncorrelated, and unintended strain in one part of the array affects other sensor readouts. This observation is particularly pronounced in the reconstruction of the shape shown in Fig. \ref{fig:extruded_shapes_results}g, where the strong central bend made it difficult to conform the board completely to the fixture surface. Although the full set of constraints still produced a good overall reconstruction with a small shape error, the slight rightward shift of the bend led to large variations in the curvature measured at the sample points. The presence of correlated error sources in the measurements suggests that the additive-white-noise analysis likely provides an upper bound on the effect on reconstruction performance, rather than an accurate prediction. The rightmost column in Fig. \ref{fig:extruded_shapes_results} shows the radiation patterns of the curved array shapes before and after appropriate phase correction is applied to each radiator, including steering of the bent array to broadside and to $\pm 30^{\circ}$. For each shape, the pattern is normalized to the maximum beam power when the array is steered to broadside, and in most shapes the power in the steered beam remains within 1.5 dB of that at broadside. While the power in the main lobe is adequately recovered after phase correction, several shapes exhibit elevated sidelobes in their steered beams. However, these are likely inherent to the steering of such curved shapes, which may result in shading, element separations greater than $\lambda/2$, and the effects of non-isotropic individual radiator patterns, rather than a result of the reconstruction error. This conclusion is supported by the results of the subsequent experiment, in which radiation patterns were synthesized using phase correction derived from two different shape-estimation methods, and produced similar radiation patterns.

\begin{figure*}
    \centering
    \includegraphics[width=0.9\linewidth]{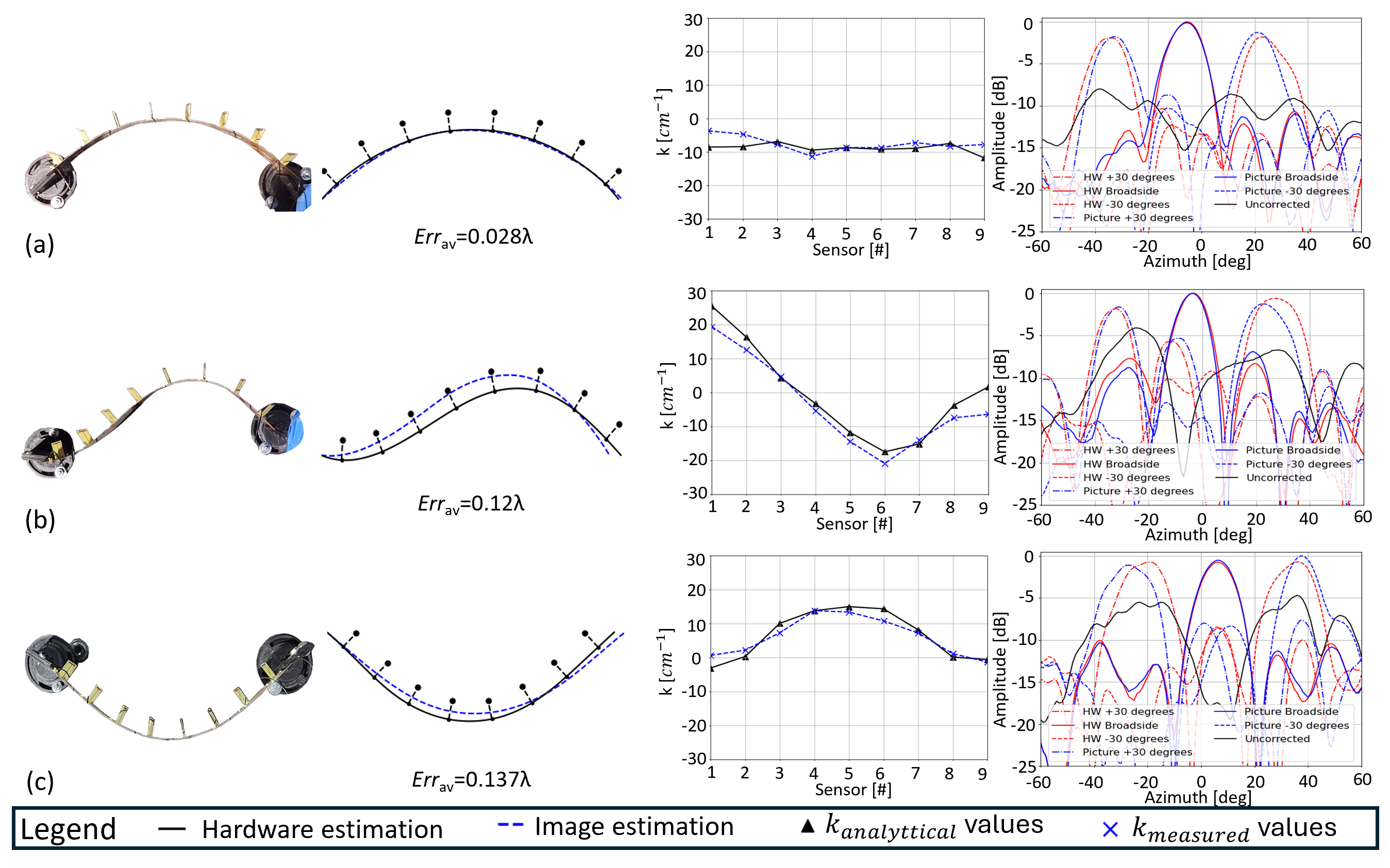}
    \caption{(Left to right) Air suspended phased array mounted on rotating clips. For each orientation, the reconstructed shape obtained from image-based analysis is compared with the hardware-estimated geometry,$k_{HW}$ compared to $k_{image}$ and the corresponding beam reconstruction and steering performance are illustrated.}
    \label{fig:air_suspended_shapes_results}
\end{figure*}

\newcommand{\rowhead}[1]{%
\textbf{\begin{tabular}[c]{@{}l@{}}#1\end{tabular}}%
}

\begin{table*}[!t]
\centering
\caption{Comparison of Reported Flexible Phased-Array Shape Sensing and Pattern-Correction Methods}
\label{tab:flexible_array_comparison}

\renewcommand{\arraystretch}{1.15}
\setlength{\tabcolsep}{3.0pt}
\scriptsize

\resizebox{\textwidth}{!}{%
\begin{tabular}{|l|c|c|c|c|c|c|c|c|}
\hline

\textbf{} &
\textbf{\shortstack{\cite{braaten2013selflex}}} &
\textbf{\shortstack{\cite{ShapeCal}}} &
\textbf{\shortstack{\cite{matan2022npj},\cite{ali2020jssc}}} &
\textbf{\shortstack{\cite{Joshi2024optical_Shapecal}}} &
\textbf{\shortstack{\cite{Zhi2025Gyroscopes}}} &
\textbf{\shortstack{\cite{Oren2026_2D}}} &
\textbf{\shortstack{This work}} \\
\hline

\rowhead{Sensing Method} &
\shortstack{Resistive\\flex sensor} &
\shortstack{Mutual\\coupling} &
\shortstack{Remote-receiver\\RF power feedback} &
\shortstack{3-D depth\\camera} &
\shortstack{6-axis\\IMUs} &
\shortstack{Mutual\\coupling} &
\shortstack{Curvature\\sensors} \\
\hline

\rowhead{Reconstruction/\\correction Method} &
\shortstack{Single-point \\deformation mapping} &
\shortstack{Spiral Match \\+ EDM optimization} &
\shortstack{Adaptive orthogonal-basis \\search} &
\shortstack{Depth-map edge \\detection + linear fitting} &
\shortstack{Orientation-based \\position reconstruction} &
\shortstack{Nonlinear least-squares (LM)} &
\shortstack{Segmented cubic-\\curve reconstruction} \\
\hline

\rowhead{RF-Independent\\Sensing} &
Yes &
No &
No &
Yes &
Yes &
No &
Yes \\
\hline

\rowhead{Requires Additional\\Hardware} &
\shortstack{Resistive\\sensors} &
-- &
\shortstack{Remote RF/\\antenna receiver} &
\shortstack{Depth/IR\\camera} &
\shortstack{IMU per\\element} &
\shortstack{2D Mutual Coupling} &
\shortstack{Strain\\gauges} \\
\hline

\rowhead{Computational\\Complexity} &
$O(1)$ &  
$O(N^2)$ &
$O(N\cdot log(P)\cdot S)^{b}$ &
-- &
$O(N)$ &  
$O(N^2)$ &
$O(N)$ \\ 
\hline

\rowhead{Demonstrated \\Geometries} &
\shortstack{Cylindrical /\\Wedge} &
\shortstack{Concave /\\convex} &
\shortstack{Planar / concave / \\convex / doubly curved} &
\shortstack{Planar / \\symmetric bend / \\asymmetric bend / \\multi-bend} &
\shortstack{Planar / \\cantilever bending / \\bowl bending} &
\shortstack{Concave/ \\convex/ \\asymmetrical multibending} &
\shortstack{Smooth arbitrary 1-D \\$R_{min}\leq3.6cm$} \\
\hline

\rowhead{Position/\\Shape Error} &
-- &
\shortstack{$\sim0.06\lambda$} &
-- &
$<1^\circ$$^{a}$ &
$<0.0225\lambda$ &
$\sim0.08\lambda$ &
$\sim0.06\lambda$ \\
\hline

\rowhead{Pattern\\Correction} &
Measured &
\shortstack{Not\\demonstrated} &
\shortstack{Measured\\RF feedback} &
Measured &
Measured &
Predicted &
Measured \\
\hline

\rowhead{Frequency} &
2.47 GHz &
\shortstack{2.5 GHz/\\10 GHz} &
10 GHz &
K-band &
2.25 GHz &
$\sim$5 GHz &
6 GHz \\
\hline

\rowhead{Array} &
$1\times4$ &
\shortstack{$1\times8$} &
\shortstack{$16\times16$ flexible\\up to $20\times20$ feedback} &
$2\times1$ tiles &
$1\times8$ &
$8\times8$ &
$1\times8$ \\
\hline

\rowhead{Demonstrated Beam\\Steering} &
$\pm30^\circ$ &
-- &
\shortstack{$\pm10^\circ$\\2-D} &
\shortstack{$110^\circ$\\total range} &
$\pm40^\circ$ &
-- &
$\pm30^\circ$ \\
\hline

\end{tabular}%
}

\vspace{1mm}

\begin{minipage}{0.98\textwidth}
\scriptsize
$^{a}$ Angular estimation error between adjacent tiles rather than
element-position reconstruction error.

$^{b}$ Complexity of computation depends on Number of elements (N) segmentation of phase states (P) and number of segmentation's of the transmitter for the calibration(S).

$^{c}$ The work in \cite{matan2022npj, ali2020jssc} is designed for static phased-arrays.

\end{minipage}

\end{table*}

\subsection{Air-suspended curved flexible array}
Once reconstruction performance was established, we conducted a second experiment to test our algorithm on naturally occurring shapes that were not defined analytically. The left column of Fig. \ref{fig:air_suspended_shapes_results} illustrates the test setup in which the flexible array was air-suspended, with two plastic clamps anchored 16 cm apart holding it at its edges. Bending was obtained by setting the angles of the clamps relative to each other, resulting in convex, concave, and doubly-curved configurations. Since the shapes were not produced analytically, we compared our hardware-based reconstruction to an image-based reconstruction produced from a top-view image of the array. The reconstruction error is quantified as the average Euclidean distance between the sensor-estimated and the image-estimated element positions. The average reconstruction error over the three inspected shapes is 9.5\%, slightly worse than that obtained for the solid surfaces in the previous experiment. The degraded accuracy might have resulted from comparison to an inaccurate image-based reconstruction due to lens aberrations, which can also be seen in the setup images, or from unaccounted vertical tilt of the air-suspended array, which may have produced a twist affecting the sensor readouts. The second-rightmost column in Fig. \ref{fig:air_suspended_shapes_results} also shows larger curvature measurement errors in the outermost sensors compared to Fig. \ref{fig:extruded_shapes_results}. This is expected, because in the air-suspended case, external force was exerted only at the clamps, leading to high tension at the array endpoints, which may not be reflected accurately in the captured image. The rightmost column compares the radiation patterns of the array at broadside and $\pm 30^{\circ}$ when the shape is reconstructed either from curvature or from image information. Notably, the broadside and first sidelobes are similar when the two different datasets were used for the reconstruction for all the shapes under test. This implies that the sidelobes in the patterns do not stem from curvature-based reconstruction errors. It can also be seen that broadside does not point exactly to $0^{\circ}$. This is because the array edges were not clipped exactly at the center of the circular clamps, which resulted in a change in the array's reference direction when rotated. Steering the array to $\pm 30^{\circ}$ shows good agreement between the two reconstruction methods, except for $-30^{\circ}$ in the shape shown in Fig. \ref{fig:air_suspended_shapes_results}c. However, given that the reference plane of the array in this test is at $\sim10^{\circ}$, the hardware reconstruction in this case is actually slightly closer to the expected result than the image-based one, emphasizing the inherent inaccuracy in treating image reconstruction as an absolute reference.
Additional RF validation of the curvature-based reconstruction approach was reported in \cite{Elgarat2026dynamic}, where radiation patterns corrected using curvature-derived geometries were compared with those corrected using independently obtained image-based geometries. The results demonstrated close agreement in main-beam direction, peak power, and half-power beam-width across three deformed configurations. Since both methods were evaluated under the same physical deformations, this agreement supports the conclusion that curvature-based reconstruction introduces a limited additional RF penalty relative to the optical reference, notwithstanding deformation-dependent electromagnetic effects common to both measurements.

Finally, Table~\ref{tab:flexible_array_comparison} compares the proposed method with mutual-coupling-based shape-reconstruction approaches \cite{austin2022jom_flex,Oren2026_2D,oren2021shapecal}, a flex-sensor-based phased-array receiver \cite{braaten2013selflex}, a tile-based flexible phased-array antenna \cite{hu2022tile_ims_manos}, a camera-based shape-reconstruction and beam-steering method \cite{Joshi2024optical_Shapecal}, and a gyroscope-based shape-reconstruction and beam-steering method \cite{Zhi2025Gyroscopes}.

\section{Conclusion}
This work extends and demonstrates a method for reconstructing the shape of flexible phased arrays from local curvature measurements using strain gauges. We developed the proposed algorithm, analyzed the sampling density required for the desired reconstruction accuracy under the physical constraints of common flexible RF materials, and examined its robustness to noise under practical measurement conditions. We also presented and experimentally validated a physical model for curvature measurement using widely available strain gauges, enabling a priori design of the sensing scheme with predictable output response. The method was demonstrated using a custom-designed flexible antenna array with embedded curvature sensors and low-cost phase shifters, enabling end-to-end recovery of beam-steering performance under shape deformation for both rigid-surface and naturally occurring free-hanging geometries. The calculated average reconstruction error of $\sim6\%$ is comparable to that of previously introduced methods, while enabling operation at significantly smaller bend radii below 3.6 cm. Since the method does not require inter-element measurements or iterative feedback, it is attractive for standalone use in physically challenging environments, for single-node sensor applications, and as a complement to other recovery methods.

\section*{Acknowledgments}
The authors would like to thank the Israeli Science Foundation (ISF) for supporting this work under grant No. 1234/24.

\bibliographystyle{IEEEtran}
\bibliography{yair_bib_file}

\end{document}